\documentclass{emulateapj}

\submitted{Astrophysical Journal}

\usepackage{amsmath}
\usepackage{txfonts}
\usepackage{graphicx}
\usepackage{multirow}
\usepackage{hyperref}
\newcommand{\be}{\begin{equation}}
\newcommand{\ee}{\end{equation}}
\newcommand{\bea}{\begin{eqnarray}}
\newcommand{\eea}{\end{eqnarray}}
\newcommand{\ha}{HI}
\newcommand{\hm}{H$_2$}
\newcommand{\sax}{S$^3$-SAX}
\newcommand{\fov}{Fo\!V}

\newcommand{\h}{h}

\newcommand{\fudge}{f_{\rm HI}}

\newcommand{\rha}{r_{\rm HI}}
\newcommand{\rhm}{r_{\rm H_2}}
\newcommand{\rhahalfmass}{\rha^{\rm half}}
\newcommand{\rhmhalfmass}{\rhm^{\rm half}}

\newcommand{\mass}{M}

\newcommand{\mha}{\mass_{\rm HI}}
\newcommand{\mhm}{\mass_{{\rm H}_2}}

\newcommand{\ms}{\mass_{\rm s}}

\newcommand{\mvir}{\mass_{\rm vir}}

\newcommand{\Sigmaha}{\Sigma_{\rm HI}}
\newcommand{\Sigmahm}{\Sigma_{\rm H_2}}

\newcommand{\Omegaha}{\Omega_{\rm HI}}
\newcommand{\Omegahm}{\Omega_{{\rm H}_2}}
\newcommand{\msun}{{\rm M}_{\odot}}

\begin{document}

\title{Evolution of the Milky Way in Semi-Analytic Models: Cold Gas at $z\!=\!3$ with ALMA and SKA}

\author{D. Obreschkow$^{1,2}$, I. Heywood$^1$, S. Rawlings$^1$}
\affil{$^1$\,Astrophysics, Department of Physics, University of Oxford, Keble Road, Oxford, OX1 3RH, UK\\
$^2$\,International Centre for Radio Astronomy Research, The University of Western Australia, 35 Stirling Hwy, Crawley, WA 6009, Australia}

\begin{abstract}
We forecast the abilities of the Atacama~Large~Millimeter/submillimeter~Array (ALMA) and the Square~Kilometer~Array (SKA) to detect CO and HI emission lines in galaxies at redshift $z=3$. A particular focus is set on Milky Way (MW) progenitors at $z=3$ for their detection within 24\,h constitutes a key science goal of ALMA. The analysis relies on a semi-analytic model, which permits the construction of a MW progenitor sample by backtracking the cosmic history of all simulated present-day galaxies similar to the real MW. Results: (i) ALMA can best observe a MW at $z=3$ by looking at CO(3--2) emission. The probability of detecting a random model MW at 3-$\sigma$ in 24\,h using $75\rm\,km\,s^{-1}$ channels is roughly 50\%, and these odds can be increased by co-adding the CO(3--2) and CO(4--3) lines. These lines fall into ALMA band 3, which therefore represents the optimal choice towards MW detections at $z=3$. (ii) Higher CO transitions contained in the ALMA bands $\geq6$ will be invisible, unless the considered MW progenitor coincidentally hosts a major starburst or an active black hole. (iii) The high-frequency array of SKA, fitted with 28.8\,GHz receivers, would be a powerful instrument for observing CO(1--0) at $z=3$, able to detect nearly all simulated MWs in 24\,h. (iv) HI detections in MWs at $z=3$ using the low-frequency array of SKA will be impossible in any reasonable observing time. (v) SKA will nonetheless be a supreme \ha~survey instrument through its enormous instantaneous field-of-view (\fov). A one year pointed HI survey with an assumed \fov~of $410\rm\,deg^2$ would reveal at least $10^5$ galaxies at $z=2.95\!-\!3.05$. (vi) If the positions and redshifts of those galaxies are known from an optical/infrared spectroscopic survey, stacking allows the detection of HI at $z=3$ in less than 24\,h.\\
\end{abstract}

\keywords{Galaxy: evolution -- galaxies: ISM -- radio lines: ISM -- cosmology: theory}

\section{Introduction}\label{section_introduction}

Detecting cold gas in ordinary distant galaxies is a paramount challenge in modern astronomy. It will be addressed by two revolutionary future telescopes, the Atacama Large Millimeter/submillimeter Array (ALMA) and the Square Kilometer Array (SKA), whose prospects have been stimulating much interest for cold gas at high redshift ($z$).

Here, ``cold gas'' refers to all gas cold enough to be neutral. Such gas consists of molecules and atoms and it dominates the interstellar medium (ISM) in galaxies, mainly in the form of hydrogen and helium with a mass ratio close to 3:1. The hydrogen is called \ha~when atomic and \hm~when molecular. Cold gas owes its astrophysical importance to its particles moving slow enough for the formation of gravitationally self-bound structures on sub-galactic scales. Cold gas thus plays a primeval role in the formation of galaxies and stars \citep{Kauffmann2006,Gunawardhana2011}. Yet, the cosmic history of these processes remains puzzling, since extrapolations from available local observations to the past are complicated by the evolving conditions of the Universe, such as density, structure, and chemical composition. Hence, it is still unclear as to when the first galaxies formed, whether their cold gas was mostly atomic or molecular, how the first stars were born, and when the ISM was sufficiently chemically enriched to form planets -- to name but a few issues calling for detailed observations of distant cold gas.

The difficulty of such observations arises from the extreme faintness of the cold gas tracers. Unless exposed to exciting radiation, cold gas only glimmers in narrow emission lines at infrared, (sub)millimeter, and radio frequencies. Molecular gas is most typically found via the $J\cdot115\rm\,GHz$ (rest-frame) lines of the rotational transitions $J\rightarrow J-1$ of the $\rm^{12}C^{16}O$ molecule (hereafter CO, \citealp[e.~g.][]{Tacconi2010,Daddi2010}). Atomic gas is detected via the ``\ha~line'' or ``21\,cm line'' at 1.420\,GHz rest-frame \citep{Zwaan2005,Martin2010}. To appreciate the difficulty of cold gas detections, note that the bolometric luminosity of a single bright star, such as Rigel ($\beta$ Orionis), is 1000-times higher than the \ha~line power of the entire Milky Way (MW). Mainly for this reason, no \ha~emission has yet been seen at $z\geqslant1$, while stellar light detections at such redshifts are now an observational standard \citep[e.g.][]{Szomoru2011,Su2011}.

The unprecedented sensitivity of ALMA and SKA in the (sub)millimeter and radio spectrum will ease the detection of CO and \ha~emission lines at high $z$. In fact, SKA was originally conceived as a pure \ha-telescope \citep{Wilkinson1991}, and the first science goal of ALMA is to ``detect spectral line emission from CO or CII in a normal galaxy like the MW at a redshift of $z=3$, in less than 24\,h of observation''\,\footnote{all science goals at http://almascience.eso.org/about-alma/full-alma}. In preparation for ALMA and SKA, predictions of their findings are needed to optimize the telescope designs, to outline initial survey strategies, and to ensure an unbiased check of our current theories against future observations.

This paper illustratively predicts the  abilities of ALMA and SKA to detect CO and \ha~emission lines from galaxies at $z=3$, based on a semi-analytic galaxy model \citep{Obreschkow2009b}. Of particular interest is the detection of lines in ``galaxies like the MW at $z=3$'' -- the first ALMA science goal. It is necessary to clarify whether this means a galaxy identical to the MW placed at a cosmological distance corresponding to $z=3$, or rather a plausible MW progenitor at $z=3$. The former interpretation is more common, but we here adopt the latter for it is perhaps more sensible in a study ultimately dedicated to the understanding of our own origins. Yet, this interpretation complicates the predictions as they require a model of the MW at a cosmic time corresponding to $z=3$, i.~e.~11 billion years back in time.

Section \ref{section_simulation} reviews the \sax~simulation\footnote{online access at http://s-cubed.physics.ox.ac.uk}. Section \ref{section_mw_evolution} studies the cosmic evolution of MW-type galaxies in \sax~with a focus on the CO and \ha~lines at $z=3$. Section \ref{section_telescope_specifications} summarizes current specifications of ALMA and SKA, based on which Sections \ref{section_detections_mw} and \ref{section_detections_general} predict the detectability of emission lines from MW-type galaxies at $z=3$ and arbitrary galaxies at $z=3$, respectively. Section \ref{section_conclusion} concludes the paper.

\section{The \sax~simulation}\label{section_simulation}

Our analysis relies on \sax~\citep{Obreschkow2009b}, a computer model of neutral atomic (\ha) and molecular (\hm) hydrogen in galaxies$^2$, which builds on the Millennium simulation \citep{Springel2005}. The Millennium simulation is a gravitational $N$-body simulation of about $10^{10}$ dark matter particles in a cubic comoving volume of $(500/h\rm\,Mpc)^3$. It models the formation of cosmic structure down to galaxy haloes as low in mass as those of the Small Magellanic Cloud (SMC), while tracking features as large as the Baryon Acoustic Oscillations (BAOs). The cosmological parameters of the Millennium simulation are $\h=0.73$, where the Hubble constant $H_0\equiv100\,h\rm\,km\,s^{-1}\,Mpc^{-1}$, $\Omega_{\rm matter}=0.25$, $\Omega_{\rm baryon}=0.045$, $\Omega_\Lambda=0.75$, $\sigma_8=0.9$.

Through a post-processing of the Millennium simulation, \citet[][see also \citealp{Croton2006}]{DeLucia2007} studied the evolution of idealized model-galaxies placed at the centers of the dark matter haloes. The global galaxy properties, such as stellar mass, cold gas mass, and morphology, were evolved according to discrete, simplistic rules. This ``semi-analytic'' processing resulted in a catalog of evolving and merging galaxies. The number of galaxies at a cosmological time of $13.7\cdot10^9\rm~yrs$ (i.~e.~today) is about $3\cdot10^7$, and each of these galaxies has a well-defined history of growing and discretely merging progenitor galaxies that have been stored in 64 discrete cosmic time steps.

\cite{Obreschkow2009b} applied an additional post-processing to the galaxies in the semi-analytic simulation by \cite{DeLucia2007} in order to subdivide their cold gas masses into \ha, \hm, and Helium. They also assigned realistic radial distributions and velocity profiles to the \ha~and \hm~components. Subsequently, \cite{Obreschkow2009e} introduced a model to assign approximate CO line luminosities to the molecular gas of each galaxy. This model relies on a single gas phase in thermal equilibrium with frequency-dependent optical depths, and it approximately accounts for the following mechanisms: (i) molecular gas is heated by starbursts, AGNs, and the redshift-dependent CMB; (ii) overlapping clouds in dense and inclined galaxies cause CO self-shielding; (iii) in compact galaxy cores molecular gas transits from a clumpy to a smooth distribution; (iv) CO-luminosities are metallicity dependent; (v) CO-luminosities are always measured relative to the redshift-dependent CMB. The integrated CO and \ha~line luminosities were further expanded into frequency-dependent profiles -- typically double-horn profiles -- by applying mass models and random galaxy-inclinations (sine-distribution). The semi-analytic galaxy model \citep{DeLucia2007} with our additional properties for \ha, \hm, and CO is called ``\sax-Box'' as a reminder that the simulated evolving galaxies are contained within the cubic volume (box) of the Millennium simulation. 

Given \sax-Box we then constructed a virtual sky \citep{Obreschkow2009f} by mapping the Cartesian coordinates (x, y, z) of the simulated galaxies onto apparent positions (RA, Dec, $z$), using the method of \cite{Blaizot2005}. Alongside this mapping, the intrinsic CO and \ha~luminosities of each galaxy were transformed into observable integrated line-fluxes. The resulting virtual sky simulation is called ``\sax-Sky'' in contrast to \sax-Box. The maximal field-of-view (\fov) of \sax-Sky depends on the selected maximal redshift $z$. At $z=3$ the \fov~is approximately $37.2\rm\,deg^2$, corresponding to the comoving surface area of $(500/h\rm\,Mpc)^2$ of the Millennium box, which is large enough to suppress significant effects of cosmic variance.

In this paper we are using both \sax-Box and \sax-Sky. \sax-Box contains the pointers needed to backtrack the cosmic evolution of MW-type galaxies to $z=3$ (Section \ref{section_mw_evolution}), while \sax-Sky provides the apparent positions and line fluxes required to study the detectability of the simulated galaxies (Sections \ref{section_detections_mw} and \ref{section_detections_general}). Throughout the whole paper, we assume that the continuum emission can be perfectly subtracted, such that the line emission can be studied independently. For other assumptions, limitations, and uncertainties of the \sax~simulation, please refer to Section 6 in \cite{Obreschkow2009b} and Section 6.2 in \cite{Obreschkow2009e}.

\section{Evolution of simulated MW-type galaxies}\label{section_mw_evolution}

We shall now investigate the cosmic evolution of the CO and \ha~line signatures of the galaxies ``like the MW'' in the \sax-Box~simulation (see Section \ref{section_simulation}).

\subsection{Definition of simulated MW-type galaxies}\label{subsection_defmw}

By definition, we call a model galaxy at $z=0$ a ``MW-type'', if its morphological type, derived from the bulge-to-disk ratio \citep[see eq.~18 in][]{Obreschkow2009b}, is Sb--Sc, and if it matches the stellar mass $\ms$, the \ha~mass $\mha$, the \hm~mass $\mhm$, the \ha~half-mass radius $\rhahalfmass$, and the \hm~half-mass radius $\rhmhalfmass$ of the MW, given in Tab.~\ref{tab_mw_coldgas}, within a factor 1.3. This factor approximately corresponds to the empirical uncertainties. According to this definition, the \sax-Box~simulation contains 1928 MW-type galaxies at $z=0$. A simulated galaxy at redshift $z>0$ is called a ``MW-type'' galaxy or a ``MW progenitor'', if, at its particular redshift, it is the most massive progenitor of a MW-type galaxy at $z=0$.

\subsection{Evolution of \ha~and \hm~in simulated MW-type galaxies}\label{subsec_hahminmw}

\sax-Box consists of 64 discrete cosmic time steps (Section \ref{section_simulation}). The cosmic evolution of any galaxy through those time steps can be extracted using a system of galaxy identifiers and progenitor-pointers that was already installed in the underlying semi-analytic galaxy model (\citealp{Croton2006}, see also \citealp{Springel2005}). We here used those pointers to follow the cosmic history of the 1928 present-day MW-type galaxies (Section \ref{subsection_defmw}).

\begin{figure}[t]
  \includegraphics[width=\columnwidth]{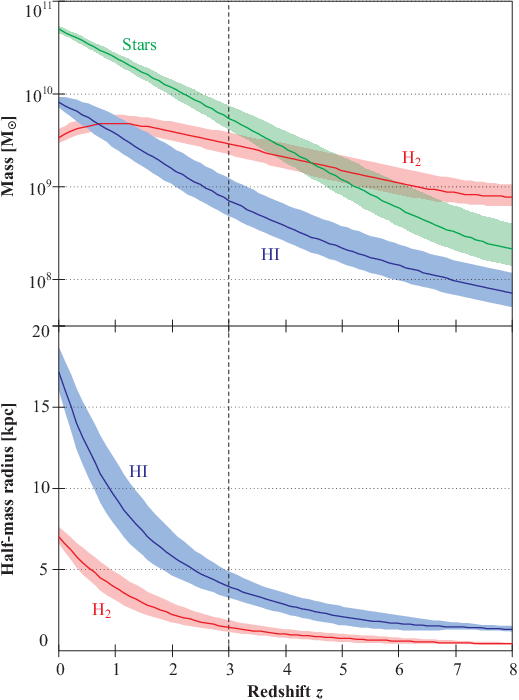}
  \caption{Cosmic evolution of the average properties of the simulated MW progenitors. The average values (solid lines) are {\it arithmetic} sample averages. The shaded regions represent the 0.5-$\sigma$ scatter around the average values in the Gaussian approximation, i.~e.~those shaded regions contain about 40\% of all events. Note that the number of galaxies, from which the averages and scatters were calculated, decreases with redshift, because the different evolution scenarios for the MW start at different cosmic times (see Section \ref{subsec_hahminmw}). Also note that this figure shows the \ha~masses without the correction introduced in Section \ref{subsec_missing_ha}.}
  \label{fig_evolution}
\end{figure}

The cosmic evolution of the sample averages of the masses and radii of the simulated MW progenitors is displayed in Fig.~\ref{fig_evolution}; and the specific values at $z=3$ have been summarized in Tab.~\ref{tab_mw_coldgas}. We emphasize that the sample size of 1928 MW-type galaxies at $z=0$ decreases monotonically with $z$, since the different evolution scenarios of the MW-type galaxies start at different initial redshifts, depending on the respective dark matter distribution. In other words, with increasing $z$, the average MW properties displayed in Fig.~\ref{fig_evolution} are increasingly biased towards evolutionary scenarios, which started particularly early in the history of the Universe. At $z=3$ this is not an issue, since in 90\% (i.~e.~1731) of all simulation scenarios the MW formed before $z=3$. However, only 39\% (i.~e.~755) of all scenarios has the MW forming before $z=7$, and only 12\% (i.~e.~234) of the scenarios before $z=8$.

Detailed physical interpretations of the cosmic evolution displayed in Fig.~\ref{fig_evolution} can be found in \citet[][evolution of the \ha~and \hm~masses]{Obreschkow2009c} and \citet[][evolution of the galaxy sizes]{Obreschkow2009d}. In brief, the radius of individual disk galaxies grows in the simulation with cosmic time approximately as $(1+z)^{-1}$, consistent with optical/infrared high-redshift observations \citep{Bouwens2004,Trujillo2006,Buitrago2008}. This size evolution is reflected in the evolution of the \ha~and \hm~radii (see Fig.~\ref{fig_evolution}, bottom panel), and it is responsible for an increase in the pressure of the interstellar medium (ISM) with $z$. By virtue of the relation between the ISM pressure and the \hm/\ha~ratio \citep[e.~g.][]{Elmegreen1993,Blitz2006,Leroy2008}, the \hm/\ha~mass ratio therefore increases with $z$. This results in a roughly constant \hm~mass for the simulated MW galaxies in the redshift range $z=0-3$ (see Fig.~\ref{fig_evolution}, top panel), while the \ha~mass varies by a factor 10 in the same redshift range (but see discussion in Section \ref{subsec_missing_ha}).

The simulated MW progenitors at $z=3$ exhibit an average gas mass fraction $(\mha+\mhm)/(\ms+\mha+\mhm)$ of 40\%, respectively 55\% when correcting the \ha~masses as described in Section \ref{subsec_missing_ha}. These values lie an order of magnitude above those found in today's massive spiral galaxies \citep{Leroy2005}, and they are in good agreement with the average gas mass fraction of 44\%, recently measured in typical massive star-forming galaxies at $z\approx2.3$ \citep{Tacconi2010}.

\begin{table}[t]
  \centering\vspace{0.2cm}
\begin{tabular}{p{3.2cm}p{1.6cm}p{1.9cm}l}
\hline \\ [-2ex] Quantity & Obs.~at $z\!=\!0$ & Sim.~at $z\!=\!3$ & Ref. \\ [0.5ex] \hline \\ [-2ex]
Virial mass $\mvir$ [$\msun$] & \mbox{$1.3\!\pm0.3\!\cdot10^{12}$} & \mbox{$2.3^{+0.9}_{-0.7}\cdot10^{11}$} & (a) \\ [0.5ex]
Stellar mass $\ms$ [$\msun$] & $5.0^{+1}_{-1}\cdot10^{10}$ & \mbox{$5.3^{+3.9}_{-2.5}\cdot10^9$} & (b) \\ [0.5ex]
\ha~mass $\mha$ [$\msun$] & $8.0^{+2}_{-2}\cdot10^9$ & \mbox{$\fudge\cdot0.7^{+1.1}_{-0.5}\cdot10^9$} & (c) \\ [0.5ex]
\hm~mass $\mhm$ [$\msun$] & $3.5^{+1}_{-1}\cdot10^9$ & \mbox{$2.8^{+1.8}_{-1.3}\cdot10^9$} & (d) \\ [0.5ex]
\ha~half-mass rad.~$\rhahalfmass$ [kpc] & $15^{+5}_{-5}$ & \mbox{$3.8^{+2.1}_{-1.6}$} & (c) \\ [0.5ex]
\hm~half-mass rad.~$\rhmhalfmass$ [kpc] & $7^{+1}_{-1}$ & \mbox{$1.4^{+0.9}_{-0.7}$} & (d) \\ [0.5ex]
\hline
\end{tabular}
\caption{Observed properties of the MW versus properties of the simulated MW-progenitors at $z=3$. The indicated ranges are 1-$\sigma$ uncertainties for the  observed values and RMS scatters around the sample averages for the simulated values. The \ha~fudge factor $\fudge$ is explained in Section \ref{subsec_missing_ha}}. The observed values have been drawn from the following references: (a) \citet{McMillan2011}, (b) \citet{Flynn2006}, (c) analytic fits to $\Sigmaha(r)$ in \citet{Kalberla2008}, (d) $\Sigmahm(r)$ in Tab.~3 in \citet{Sanders1984}.
\label{tab_mw_coldgas}
\end{table}

\subsection{CO and \ha~lines of simulated MW progenitors at $z\!=\!3$}\label{subsection_hacoz3}

\begin{figure}
  \includegraphics[width=0.99\columnwidth]{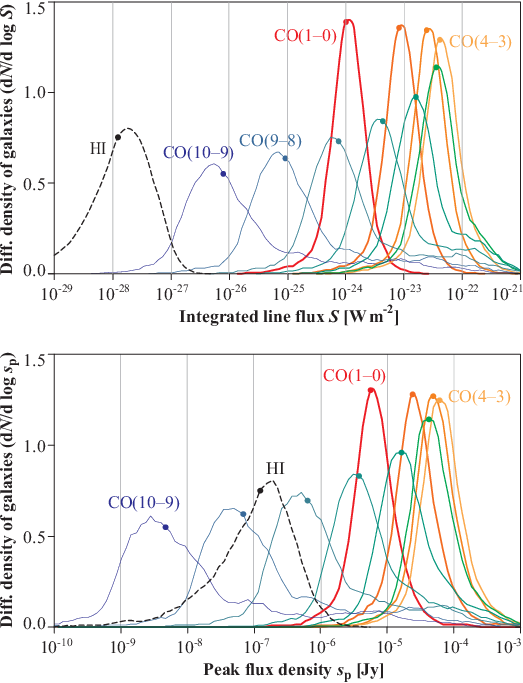}
  \caption{Normalized sample distribution of the frequency-integrated line fluxes $S$ (top) and peak flux densities $s_{\rm p}$ (bottom) for the 1731 simulated MW galaxies at $z=3$. Dots represent median values. Note that the \ha~fluxes shown here are drawn from \sax~without the correction introduced in Section \ref{subsec_missing_ha}. These data are accessible online$^{\,3}$.}
  \label{fig_co_lum_distribution}
\end{figure}

Fig.~\ref{fig_co_lum_distribution} (data available online\footnote{http://s-cubed.physics.ox.ac.uk/downloads/mw-at-z3.xls}) displays the emission line characteristics of our sample of 1731 simulated MW progenitors  at $z=3$. The upper panel shows the sample distributions of the frequency-integrated line fluxes $S$, while the lower panel represents the distributions of the peak flux densities $s_{\rm p}$. Using eq.~(A11) in \cite{Obreschkow2009e} the frequency-integrated fluxes $S$ (here in units of $\rm W\,m^{-2}$) can be converted into velocity-integrated fluxes (e.~g.~in units of $\rm Jy\,km\,s^{-1}$).

The first conclusion from Fig.~\ref{fig_co_lum_distribution} is that the sample of simulated MW progenitors covers a wide range of fluxes for each individual emission line. In fact, the root-mean-square (RMS) scatter of the line fluxes varies between 0.5 and 1 dex for the different lines. For the CO transitions up to CO(5--4) the sample distributions are roughly Gaussian in log-space, reflecting the underlying sample scatter in the \hm~mass. However, the higher-order CO transitions are skewed towards the high-flux end in the distribution. For example, the highest CO(10--9) fluxes in the sample lie nearly five orders of magnitude above the sample median. This non-Gaussian excess of high fluxes for the higher-order CO transitions reflects the relatively rare cases where the molecular gas is heated by a massive starburst or an AGN. In fact, all simulated MW scenarios occasionally undergo starbursts and phases of intense black hole accretion, but at any given cosmic time, such as the time corresponding to $z=3$, only a minor fraction ($\approx5\%$) of all galaxies in the sample is subjected to such an exceptional source of heat. We also note that the sample distribution of \ha~fluxes is the only distribution with a non-Gaussian excess in the low flux regime. This feature is again attributed to occasional black hole activity, which, in the semi-analytic setup \citep{Croton2006}, results in a suppression of the cooling flow. Due to the large scatter and non-Gaussianity of the flux distribution in the sample, ``average line fluxes'' can be ambiguous or at worst meaningless. For example the average integrated CO(10--9) flux lies three orders of magnitude above the most probable integrated CO(10--9) flux. For this reason, we shall restrict our considerations to median values, where necessary. Those values have been marked as dots in Fig.~\ref{fig_co_lum_distribution} and are listed in Tab.~\ref{tab_mw_detection} (columns 16, 17) in Section \ref{section_detections_mw}.

\begin{figure}
  \includegraphics[width=\columnwidth]{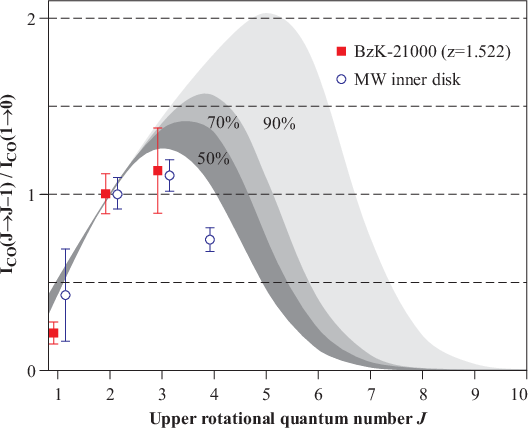}
  \caption{Comparison of simulated CO-SEDs in MWs at $z=3$ (shading) with observed data points of ``normal'' star forming galaxies: the inner MW disk \citep[circles,][]{Fixsen1999} and the massive galaxy BzK-21000 at $z=1.522$ \citep[squares,][]{Aravena2010,Dannerbauer2009}. All SEDs are expressed in velocity-integrated fluxes, normalized to CO(2--1). The shadings respectively contain 50\% of the simulated SEDs (dark gray), 70\% (dark and mid-tone gray), and 90\% (all gray tones).}
  \label{fig_cosed}
\end{figure}

How do the simulated CO Spectral Energy Distributions (CO-SEDs) of our model MWs at $z=3$ compare to real data? Recent observations of CO-SEDs in massive disk-like galaxies at $z\approx1.5$ found that those systems yield CO-SEDs similar to that of the MW, in contrast to the highly excited CO-SEDs typically observed in high-$z$ submillimeter galaxies \citep{Dannerbauer2009}. Fig.~\ref{fig_cosed} demonstrates that these new data are roughly in line with the simulated MW progenitors, 50\% of which yield CO-SEDs only slightly more excited than the inner MW disk. These low-excitation MW progenitors dominate the predictions for ALMA and SKA in Sections \ref{section_detections_mw} and \ref{section_detections_general}. However, the simulation also predicts the existence of a minority of highly excited CO-SEDs (light gray in Fig.~\ref{fig_cosed}) corresponding to the simulated MWs that underwent a starburst and/or an AGN at $z=3$.

\subsection{The missing \ha~mass problem}\label{subsec_missing_ha}
As discussed earlier \citep{Obreschkow2009c}, the \sax~model misses a significant fraction of \ha~at $z>0$ compared to inferences from damped Lyman alpha systems (DLAs). At $z=3$, the global space density of \ha~in \sax~lies a factor 5 below the DLA data. Recent efforts to understand this difference \citep{Lagos2011} claim that it can be widely explained by the limited mass resolution of the Millennium simulation, which defines a lower mass limit of $\mvir\approx10^{10}\msun$ for the semi-analytic model of \cite{DeLucia2007}.  Based on a Monte-Carlo extrapolation to smaller galaxies \cite{Lagos2011} find that most of the \ha~gas at redshift $z=3$ resides in galaxies that cannot be resolved in the Millennium simulation. This explanation of the missing \ha~mass in \sax~is further supported by high-resolution smoothed particle simulations of galaxy formation \citep{Pontzen2008}, which suggest that most DLAs are associated with small halo masses of $\mvir=10^9-10^{11}\msun$. While very plausible, these results remain uncertain because the \hm/\ha~ratio of small high-$z$ galaxies is poorly understood, in particular because their geometry is likely to deviate significantly from flat disks. The cosmic evolution of metallicity and velocity dispersion \citep{ForsterSchreiber2006} adds to this uncertainty.

To address the many systematic uncertainties regarding \ha~at $z=3$, we shall here consider two models: the raw \sax~model, which seems to underestimate the space density $\Omega_{\rm HI}(z=3)$; and a heuristic correction of the \sax~model, where all simulated \ha~masses are multiplied by a fudge factor $\fudge=5$, matching $\Omega_{\rm HI}(z=3)$ inferred from DLAs. Physically, $\fudge$ can be interpreted a correction containing the non-resolved satellites, as well as an \ha-rich non-disk component around each galaxy. All the \ha~detection predictions in Sections \ref{section_detections_mw} and \ref{section_detections_general} are provided for both the raw \sax~model ($\fudge=1$) and the corrected one ($\fudge=5$).

\begin{table*}[t]
\centering
\begin{tabular}{lcccccc}
\hline \\ [-2.0ex]
  & \mbox{~~~~~~~~~ALMA~~~~~~~~~} & \mbox{~~SKA$_1$-LF~~} & \mbox{~~SKA$_2$-LF~~} & \mbox{~~SKA$_2$-MF~~} &\mbox{~~SKA$_1$-HF~~} & \mbox{~~SKA$_2$-HF~~} \\[0.5ex]
\hline \\ [-2.0ex]
\mbox{Receiver type} & \mbox{SFD} & \mbox{AAS} & \mbox{AAS} & \mbox{AAS} & \mbox{SFD} & \mbox{SFD}\\[0.5ex]
\mbox{Diameter of dishes/stations $D\rm~[m]$} & \mbox{12} & \mbox{180} & \mbox{180} & \mbox{56} & \mbox{15} & \mbox{15}\\[0.5ex]
\mbox{Number of dishes/stations $N_{\rm units}$} & \mbox{50} & \mbox{50} & \mbox{250} & \mbox{250} & \mbox{125$^\ast$} & \mbox{1250$^\ast$}\\[0.5ex]
\mbox{Number of instantaneous beams $N_{\rm beams}$} & \mbox{1} & \mbox{480} & \mbox{4800} & \mbox{4800} & \mbox{1} & \mbox{1}\\[0.5ex]
\mbox{RMS surface error of dishes $\delta\rm~[mm]$} & \mbox{0.01} & \mbox{--} & \mbox{--} & \mbox{--} & \mbox{0.5} & \mbox{0.5}\\[0.5ex]
\mbox{Geometry factor $\epsilon_{\rm g}$} & \mbox{1} & \mbox{eq.~(\ref{eq_aeff})} & \mbox{eq.~(\ref{eq_aeff})} & \mbox{eq.~(\ref{eq_aeff})} & \mbox{1} & \mbox{1}\\[0.5ex]
\mbox{Correlator quantization efficiency $\epsilon_{\rm q}$} & \mbox{0.95} & \mbox{0.95} & \mbox{0.95} & \mbox{0.95} & \mbox{0.95} & \mbox{0.95}\\[0.5ex]
\mbox{Array efficiency $\epsilon_{\rm x}$} & \mbox{0.90} & \mbox{0.90} & \mbox{0.90} & \mbox{0.90} & \mbox{0.90} & \mbox{0.90}\\[0.5ex]
\mbox{Antenna efficiency $\epsilon_{\rm a}$} & \mbox{eq.~(\ref{eq_ruze})} & \mbox{0.90} & \mbox{0.90} & \mbox{0.90} & \mbox{eq.~(\ref{eq_ruze})} & \mbox{eq.~(\ref{eq_ruze})}\\[0.5ex]
\mbox{Nyquist-sampling frequency $\nu_0\rm~[MHz]$} & \mbox{--} & \mbox{115} & \mbox{115} & \mbox{800} & \mbox{--} & \mbox{--}\\[0.5ex]
\mbox{RMS baseline of compact configuration $B_{\rm RMS}\rm~[km]$} & \mbox{0.08} & \mbox{100} & \mbox{100} & \mbox{100} & \mbox{0.5$^\ast$} & \mbox{0.5$^\ast$}\\[0.5ex]
\mbox{Receiver temperature $T_{\rm rec}\rm~[K]$} & \mbox{Tab.~\ref{tab_sensitivities}, col.~(\ref{col_trec})} & \mbox{150} & \mbox{150} & \mbox{50} & \mbox{30} & \mbox{30}\\[0.5ex]
\mbox{Sky temperature $T_{\rm sky}\rm~[K]$} & \mbox{Tab.~\ref{tab_sensitivities}, col.~(\ref{col_tsky})} & \mbox{eq.~(\ref{eq_tsky_HI})} & \mbox{eq.~(\ref{eq_tsky_HI})} & \mbox{eq.~(\ref{eq_tsky_HI})} & \mbox{2.7} & \mbox{2.7}\\[0.5ex]
\hline
\end{tabular}
   \caption{General specifications of ALMA and SKA. AAS stands for ``aperture array station''; SFD for ``single-feed dish''. The values marked with $^\ast$ correspond to the CORE of SKA-HF. The full SKA-HF has twice the number of dishes and long baselines with $B_{\rm RMS}\approx100\rm~km$. Core-only CO(1--0)~observations are considered in this paper, since the high spatial resolution (\i.~e.~low surface brightness sensitivity) of the full array would yield far less individual CO(1--0) detections at $z=3$ (see Section \ref{subsection_spatial_resolution}).}\vspace{0.5cm}
   \label{tab_telescope_properties}
\end{table*}

\section{Specifications of ALMA and SKA}\label{section_telescope_specifications}
This section outlines the provisional specifications of ALMA and SKA, needed for the predictions in Sections \ref{section_detections_mw} and \ref{section_detections_general}. Over the following paragraphs we present the physical concepts and assumptions behind the fundamental telescope parameters listed in Tab.~\ref{tab_telescope_properties} and the derived emission line specific parameters listed in Tab.~\ref{tab_sensitivities}.

\subsection{Brief overview of ALMA}
ALMA is a reconfigurable array of 50 steerable single-feed dishes (SFDs). By exchanging the receivers, 10 different frequency bands, here called ALMA-1 to ALMA-10, can be reached. They collectively cover the whole atmospherically transparent parts of the spectrum between 31.3\,GHz and 950\,GHz. The window between 31.3\,GHz and 84\,GHz (ALMA-1 and ALMA-2) remains subject to future receiver development, and the availability of the window between 163\,GHz and 211\,GHz (ALMA-5) is still uncertain, potentially impeding CO(6--5) and CO(7--6) observations at $z=3$. The general ALMA specifications in Tab.~\ref{tab_telescope_properties} summarize the current online specifications\footnote{http://science.nrao.edu/alma/specifications.shtml}. Before the completion of the array by 2013, the Large Millimeter Telescope (LMT) yields competitive sensitivities in the ALMA spectrum from $\sim80$ to $350$\,GHz.

\begin{figure}[t]
  \includegraphics[width=\columnwidth]{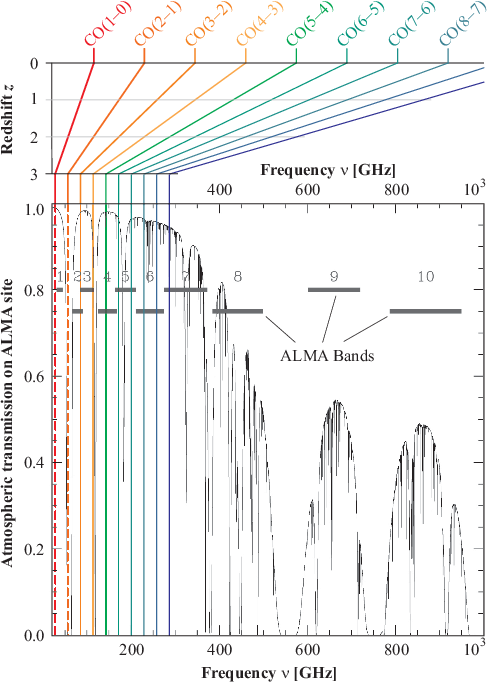}
  \caption{CO emission line frequencies at $z=3$ compared to the atmospheric transmissivity at the ALMA site (transmission function reprinted from https://almascience.nrao.edu/about-alma/alma-site). All considered CO transitions except for CO(2--1) exhibit a high transmissivity, i.~e.~low atmospheric absorption. Note that the availability of bands 5 [CO(6--5) and CO(7--6)] is not yet confirmed.\\}
  \label{fig_atmosphere}
\end{figure}

All redshifted CO emission lines at $z=3$ considered in this paper are covered by the ALMA-bands except for the CO(1--0) line at 28.8\,GHz, which is too low in frequency, and the CO(2--1) line at 57.6\,GHz, which lies at the center of a major oxygen absorption band. For the remaining CO lines, the atmospheric transmissivity is remarkably high, such as illustrated in Fig.~\ref{fig_atmosphere}) for a low precipitable water vapour (PWV) of 0.5\,mm. In this paper, we adopt a slightly less optimistic value of PWV=1.0\,mm, which is realistic in the sense that lower (i.~e.~better) PWV-values have been measured over more than 50\% of the time over a year at the ALMA site\footnote{http://www.apex-telescope.org/sites/chajnantor/atmosphere}.

\subsection{Brief overview of SKA}
SKA is only approximately specified and its concept might still change. Here we assume that SKA will be composed of three independent arrays with specifications synthesizing those described by \cite{Dewdney2010}, \cite{Garrett2010}, and \cite{Schilizzi2007}: the low-frequency array ``SKA-LF'', operating at 70\,MHz--450\,MHz (\ha~at $z=19.3\!-\!2.2$); the mid-frequency array ``SKA-MF'' at 400\,MHz--1.4\,GHz (\ha~at $z=2.6\!-\!0$); and the high-frequency array ``SKA-HF'' at 1\,GHz--30\,GHz\,\footnote{The SKA design process currently uses 10 GHz as the required frequency upper limit. However, it also targets 0.5 mm RMS surface accuracy to ensure high dynamic range imaging \citep{Garrett2010}. Hence, the Ruze-equation (eq.~\ref{eq_ruze}) shows that the dishes should have good efficiencies up to 30 GHz.} (e.~g.~CO(1--0) line at $z>2.8$). The latter is a fixed array of steerable SFDs, whereas SKA-LF and SKA-MF are fixed arrays of circular aperture array stations (AASs) -- a modern concept with no moving parts, currently realized in the European Low-Frequency Array (LOFAR). In this paper, we only consider SKA-LF (for the \ha~line at $z=3$) and SKA-HF (for the CO(1--0) line at $z=3$); but see the discussion in Section \ref{subsec_discussion_ska} regarding potential uses of SKA-MF for \ha~detections at $z\approx3$.

SKA will be deployed in two phases referred to as SKA$_1$ and SKA$_2$. The main differences between these phases are summarized in Tab.~\ref{tab_telescope_properties}. In particular, SKA$_2$-HF is assumed to have 10-times more SFDs than SKA$_1$-HF, and SKA$_2$-LF is assumed to have 5-times more AASs than SKA$_1$-LF. Also note that in the current design, the mid-frequency array SKA-MF will only be added in phase-2. The completion of SKA$_2$ is not expected before 2022, i.~e.~around a decade after the completion of ALMA. Until then, several other telescopes and networks, summarized in Section 3 of \cite{Rawlings2011}, will serve as technological and scientific SKA-pathfinders: the Australian SKA Pathfinder (ASKAP), the South African SKA Pathfinder (MeerKAT), the Westerbork Synthesis Radio Telescope (WSRT) upgraded with the phase array feed (APERTIF), the Murchison Widefield Array (MWA), the Low-Frequency~Array (LOFAR), the upgraded Multi-Element Radio Linked Interferometer Network (e-MERLIN), the electronic European VLBI Network (e-EVN), the European Pulsar Timing Array (EPTA), and the Five hundred meter Aperture Spherical Telescope (FAST). Also the following instruments will help preparing the way towards SKA: the Hydrogen Epoch of Reionization Array (HERA), the Extended Very Large Array (EVLA), the Giant Meterwave Radio Telescope (GMRT) with its new software correlator, and the Arecibo telescope via the Arecibo Legacy Fast ALFA Survey (ALFALFA).

\subsection{Point-source sensitivity}\label{subsection_sensitivity}
The RMS noise $\sigma$ (units proportional to Jy) of arrays of SFDs and AASs is approximated by \citep[e.~g.][]{DeBreuck2005}
\be\label{eq_sigma}
\sigma = \frac{2\,k\,T_{\rm sys}}{A\,\epsilon_{\rm g}\epsilon_{\rm q}\epsilon_{\rm x}\epsilon_{\rm a}\sqrt{n_{\rm p}\Delta\nu\Delta t}}\,,
\ee
where $T_{\rm sys}=T_{\rm rec}+T_{\rm sky}$ is the system temperature that describes the noise of the receiver ($T_{\rm rec}$) and the sky ($T_{\rm sky}$) in the black body approximation; $k$ is the Boltzmann constant; $n_{\rm p}$ is the number of co-added polarizations (i.~e.~$n_{\rm p}=2$ if information on polarization is irrelevant), $\Delta\nu$ is the width of the frequency channels; $\Delta t$ is the integrated observing time; and $A$ is the physical surface area, that is literally the summed surface area of all SFDs or AASs.

The dimensionless parameters $\epsilon_{\rm g}$, $\epsilon_{\rm q}$, $\epsilon_{\rm x}$, $\epsilon_{\rm a}$ in eq.~(\ref{eq_sigma}) are efficiency terms, all between 0 and 1, associated with the dominant causes of sensitivity loss. The dimensionless ``geometry factor'' $\epsilon_{\rm g}$ corrects $A$ for geometrical projection effects and under-sampling effects in the case of AASs, while $\epsilon_{\rm g}=1$ for dishes. Explicitly,
\be\label{eq_aeff}
\epsilon_{\rm g}=\min\left[\cos\theta,\frac{\lambda_{\rm obs}}{\lambda_0}\cos\theta,\left(\frac{\lambda_{\rm obs}}{\lambda_0}\right)^2\right]~~\rm for~AASs,
\ee
where $\theta$ is the zenith-angle of the point-source, $\lambda_{\rm obs}$ is the observing wavelength, and $\lambda_0\equiv c/\nu_0$ is the Nyquist-sampling wavelength of the array. $\epsilon_{\rm g}\equiv\epsilon_{\rm g1}\cdot\epsilon_{\rm g2}$ is the product of two interpretable factors: $\epsilon_{\rm g1}\equiv\cos\theta$ corrects the collecting area for the linear projection of the wavefront onto the horizontal array; $\epsilon_{\rm g2}\equiv\min[1,\lambda_{\rm obs}/\lambda_0]\min[1,\lambda_{\rm obs}/(\lambda_0\cos\theta)]$ ensures that the sensitivity is correctly reduced ($\sigma$ is increased) in the case of a sparse (i.~e.~sub-Nyquist) sampling of the wavefront. The ``correlator quantization efficiency'' $\epsilon_{\rm q}$ measures the noise level output by the correlator (software or hardware) compared to that of an ideal correlator; we here assume high-end correlators with $\epsilon_{\rm q}=0.95$. The ``array efficiency'' $\epsilon_{\rm x}$ measures the losses due to the time needed to reconfigure the array and due the differential weighting applied to the visibilities (i.~e.~tapering of the $(u,v)$-plane, see \citealp{Holdaway1998,Yun1999}). Here we adopt the optimistic value of $\epsilon_{\rm x}=0.9$ as we only consider fixed array configurations (the most compact configuration for ALMA and the permanent configuration for SKA), working in a low resolution mode close to the natural weighting of the visibilities (low tapering). Finally, the ``antenna efficiency'' (sometimes called ``aperture efficiency'') $\epsilon_{\rm a}$ is the fraction of the electromagnetic energy transmitted from the collector to the receiver. For dishes, $\epsilon_{\rm a}$ can be approximated by the Ruze-equation \citep{Ruze1952}
\be\label{eq_ruze}
\epsilon_{\rm a} = \epsilon_0\exp\left[-\left(\frac{4\pi\delta}{\lambda_{\rm obs}}\right)^2\right],
\ee
where $\epsilon_0\approx0.8$ is the long wavelength maximum efficiency, $\delta$ is the RMS value of the surface error distribution (values given in Tab.~\ref{tab_telescope_properties}), and $\lambda_{\rm obs}$ is the observed wavelength.

We note that $\sigma$ in eq.~(\ref{eq_sigma}) is the RMS of the random component of the observing noise (in units proportional to Jy), measured per frequency channel of width $\Delta\nu$ and per synthesized beam, i.~e.~per pixel of the synthesized sky image. Hence $\sigma$ is the per-channel-noise of any source small enough to be fully contained within the synthesized beam. This feature applies in particular and always to point-sources. Therefore the sensitivity defined as the inverse of the noise $\sigma$ is often referred to as ``point-source sensitivity''. For resolved sources, the sensitivity deteriorates, such as detailed in Section \ref{subsection_spatial_resolution}.

\subsection{Instantaneous field-of-view}\label{subsection_fov}
The full-width-half-maximum (FWHM) of the primary beam of an AAS or a SFD with diameter $D$ is approximately
\be\label{eq_fwhm}
{\rm FWHM}\approx1.22\,\lambda_{\rm obs}/D,
\ee
and therefore the field-of-view (\fov)~of the primary beam is
\be\label{eq_beamfov}
{\rm \fov_{\rm beam}}\approx1.17\,(\lambda_{\rm obs}/D)^2.
\ee
AASs constantly collect electro-magnetic radiation from a large fraction of the hemisphere, typically covering about $10^4\rm\,deg^2$ (half the hemisphere). The beams are ``formed'' through digital processing, and in principle the number of instantaneous beams $N_{\rm beams}$ is only limited by the processing power. Given $N_{\rm beams}$, the instantaneous \fov~of AASs reads
\be\label{eq_instfov}
{\rm \fov}=N_{\rm beams}\,{\rm \fov_{\rm beam}}.
\ee
Note that we here assume that SKA-LF only uses digital beam forming. By contrast, the ``pathfinder'' instruments MWA and LOFAR also use analogue beam formers that cut down the \fov.

\begin{table*}[t]
\centering
\begin{tabular}{ccccccccccccccc}
\hline \\ [-2.0ex]
Emission & Telescope & \multicolumn{2}{c}{Frequency [GHz]} & $\Delta\nu$ & $~T_{\rm rec}~$ & $~T_{\rm sky}~$ & $~T_{\rm sys}~$ & $\epsilon_{\rm g}$ & $\epsilon_{\rm q}$ & $\epsilon_{\rm x}$ & $\epsilon_{\rm a}$ & RMS noise & Inst.~\fov~ & Resolution\\
 line & and band & $z=0$ & $z=3$ & [MHz] & [K] & [K] & [K] & [-] & [-] & [-] & [-] & $\rm[mJy\sqrt{\rm min}]$ & $\rm[deg^2]$ & $\gamma~\rm['']$ \\
 (1) & (2) & (3) & (4) & (5) & (6) & (7) & (8) & (9) & (10) & (11) & (12) & (13) & (14) & (15) \\ [0.5ex]
\hline \\ [-2.0ex]
\mbox{HI} & \mbox{SKA$_1$-LF} & \mbox{$1.420$} & \mbox{$0.3551$} & \mbox{$0.089$} & \mbox{150} & \mbox{39} & \mbox{189} & \mbox{$0.10$} & \mbox{$0.95$} & \mbox{$0.90$} & \mbox{$0.90$} & \mbox{$1.6$} & \mbox{$41$} & \mbox{$2.1$} \\[0.5ex]
\mbox{HI} & \mbox{SKA$_2$-LF} & \mbox{$1.420$} & \mbox{$0.3551$} & \mbox{$0.089$} & \mbox{150} & \mbox{39} & \mbox{189} & \mbox{$0.10$} & \mbox{$0.95$} & \mbox{$0.90$} & \mbox{$0.90$} & \mbox{$0.31$} & \mbox{$410$} & \mbox{$2.1$} \\[0.5ex]
\mbox{CO(1--0)} & \mbox{SKA$_1$-HF} & \mbox{$115.3$} & \mbox{$28.82$} & \mbox{$7.2$} & \mbox{30} & \mbox{3} & \mbox{33} & \mbox{$1.0$} & \mbox{$0.95$} & \mbox{$0.90$} & \mbox{$0.56$} & \mbox{$0.29$} & \mbox{$1.9\!\cdot\! 10^{-3}$} & \mbox{$5.2$} \\[0.5ex]
\mbox{CO(1--0)} & \mbox{SKA$_2$-HF} & \mbox{$115.3$} & \mbox{$28.82$} & \mbox{$7.2$} & \mbox{30} & \mbox{3} & \mbox{33} & \mbox{$1.0$} & \mbox{$0.95$} & \mbox{$0.90$} & \mbox{$0.56$} & \mbox{$0.029$} & \mbox{$1.9\!\cdot\! 10^{-3}$} & \mbox{$5.2$} \\[0.5ex]
\mbox{CO(3--2)} & \mbox{ALMA-3} & \mbox{$345.8$} & \mbox{$86.45$} & \mbox{$22$} & \mbox{37} & \mbox{8} & \mbox{45} & \mbox{$1.0$} & \mbox{$0.95$} & \mbox{$0.90$} & \mbox{$0.80$} & \mbox{$0.63$} & \mbox{$3.2\!\cdot\! 10^{-4}$} & \mbox{$11$} \\[0.5ex]
\mbox{CO(4--3)} & \mbox{ALMA-3} & \mbox{$461.1$} & \mbox{$115.3$} & \mbox{$29$} & \mbox{37} & \mbox{48} & \mbox{85} & \mbox{$1.0$} & \mbox{$0.95$} & \mbox{$0.90$} & \mbox{$0.80$} & \mbox{$1.0$} & \mbox{$1.8\!\cdot\! 10^{-4}$} & \mbox{$8.2$} \\[0.5ex]
\mbox{CO(5--4)} & \mbox{ALMA-4} & \mbox{$576.4$} & \mbox{$144.1$} & \mbox{$36$} & \mbox{51} & \mbox{9} & \mbox{60} & \mbox{$1.0$} & \mbox{$0.95$} & \mbox{$0.90$} & \mbox{$0.80$} & \mbox{$0.65$} & \mbox{$1.2\!\cdot\! 10^{-4}$} & \mbox{$6.5$} \\[0.5ex]
\mbox{CO(6--5)} & \mbox{ALMA-5} & \mbox{$691.6$} & \mbox{$172.9$} & \mbox{$43$} & \mbox{65} & \mbox{19} & \mbox{84} & \mbox{$1.0$} & \mbox{$0.95$} & \mbox{$0.90$} & \mbox{$0.80$} & \mbox{$0.84$} & \mbox{$8.0\!\cdot\! 10^{-5}$} & \mbox{$5.5$} \\[0.5ex]
\mbox{CO(7--6)} & \mbox{ALMA-5} & \mbox{$806.9$} & \mbox{$201.7$} & \mbox{$50$} & \mbox{65} & \mbox{17} & \mbox{82} & \mbox{$1.0$} & \mbox{$0.95$} & \mbox{$0.90$} & \mbox{$0.79$} & \mbox{$0.75$} & \mbox{$5.9\!\cdot\! 10^{-5}$} & \mbox{$4.7$} \\[0.5ex]
\mbox{CO(8--7)} & \mbox{ALMA-6} & \mbox{$922.2$} & \mbox{$230.5$} & \mbox{$58$} & \mbox{83} & \mbox{16} & \mbox{99} & \mbox{$1.0$} & \mbox{$0.95$} & \mbox{$0.90$} & \mbox{$0.79$} & \mbox{$0.86$} & \mbox{$4.5\!\cdot\! 10^{-5}$} & \mbox{$4.1$} \\[0.5ex]
\mbox{CO(9--8)} & \mbox{ALMA-6} & \mbox{$1037$} & \mbox{$259.4$} & \mbox{$65$} & \mbox{83} & \mbox{19} & \mbox{102} & \mbox{$1.0$} & \mbox{$0.95$} & \mbox{$0.90$} & \mbox{$0.79$} & \mbox{$0.83$} & \mbox{$3.6\!\cdot\! 10^{-5}$} & \mbox{$3.6$} \\[0.5ex]
\mbox{CO(10--9)} & \mbox{ALMA-7} & \mbox{$1153$} & \mbox{$288.2$} & \mbox{$72$} & \mbox{147} & \mbox{24} & \mbox{171} & \mbox{$1.0$} & \mbox{$0.95$} & \mbox{$0.90$} & \mbox{$0.79$} & \mbox{$1.3$} & \mbox{$2.9\!\cdot\! 10^{-5}$} & \mbox{$3.3$} \\[0.5ex]
\hline
\end{tabular}
   \caption{Considered emission lines of CO and \ha~with the corresponding telescope properties of ALMA and SKA. Detailed descriptions of each column are provided in Section \ref{subsection_performace}.}
   \label{tab_sensitivities}
\end{table*}

\subsection{Spatial resolution and sensitivity for resolved sources}\label{subsection_spatial_resolution}
The spatial resolution $\gamma$ of telescope arrays corresponds to the beam width of the full array, which is also referred to as the ``synthesized beam'' to be distinguished from the ``beam'' of individual SFDs and AASs. We here approximate this resolution by substituting $D$ in eq.~(\ref{eq_fwhm}) for the RMS length of the baselines $B_{\rm RMS}$,
\be\label{eq_gamma}
	\gamma = 1.22\,\lambda_{\rm obs}/B_{\rm RMS},
\ee
although the precise resolution depends on the full baseline pattern, on the sky coordinates, and on the post-processing (i.~e.~weighting of the baselines). To check the resolution implied by eq.~(\ref{eq_gamma}) for SKA-LF, we explicitly computed the point spread function (PSF) for a generic configuration of fifty AASs (as illustrated in Fig.~3 of \citealp{Dewdney2010}; station positions supplied by R.~Millenaar, priv.~comm.), centered about (lon $72^{\circ}$, lat $-30^{\circ}$). The simulation consists of a complete track at 385 MHz, observing a source at a favorable declination of $-60^{\circ}$, with flagging applied to scans below $30^{\circ}$ elevation. The naturally-weighted restoring beam for such an observation corresponds to a resolution of $2.8''\times1.9''$ at a position angle of $95^{\circ}$, in close agreement with our derived angular resolution of $2.1''$ (Tab.~\ref{tab_sensitivities}, col.~\ref{col_spatial_resolution}) calculated via eq.~(\ref{eq_gamma}).

In this paper, we focus on the detectability of individual galaxies. In order to obtain the highest sensitivity, it is therefore desirable to chose the lowest possible spatial resolution. For ALMA, this is achieved by selecting the most compact array configuration ($B_{\rm RMS}=0.08\rm~km$). This yields spatial resolutions between 3.3'' (CO(10--9) at $z=3$) and 11'' (CO(3--2) at $z=3$), much larger than the average apparent \hm~half-mass diameter of the MW disks at $z=3$ of about $0.4''$, as derived from $\rhmhalfmass$ in Tab.~\ref{tab_mw_coldgas} using an angular diameter distance of $1606\rm~Mpc$. No MW progenitor will hence be resolved using the compact ALMA configuration. For SKA ($B_{\rm RMS}\approx100\rm~km$) the situation is more subtle, since this instrument is not reconfigurable. In the case of \ha~at $z=3$, the resolution of $2.1''$ still exceeds the average \ha~half-mass diameter of the MW disks at $z=3$ of about $1''$. No MW progenitor will hence be resolved. However, in the case of CO(1--0) at $z=3$ imaged with SKA-HF, the resolution becomes as good as $0.03''$, such that a typical MW progenitor will be resolved in roughly 100 pixels (depending on inclination and evolution scenario). The associated order-of-magnitude loss in surface brightness sensitivity implies that many more sources will be picked up in core-only observations, where half of SKA's collecting area is sacrificed to the benefit of having all antennas within a core of $0.5\rm~km$ radius. In this case the resolution drops to $5.2''$ and none of the MW at $z=3$ will be resolved in CO(1--0). In this paper we therefore assume that only the core of SKA-HF is used.

Given these assumptions, none of the MW progenitors at $z=3$ will be resolved in \ha~or CO emission. However, other galaxies $z=3$, larger than the MW progenitors, may still be resolved. In this case, the sensitivity will be reduced relative to the point-source sensitivity of eq.~(\ref{eq_sigma}). In fact, if every pixel has an RMS noise level $\sigma$ defined by eq.~(\ref{eq_sigma}), then a source extended over $m>1$ pixels will be subjected to a Jy-noise equal to $\sigma'=\sigma\sqrt{m}$. Since real sources are not homogeneous, but rather exponentially fading disks, the definition of $m$ is not obvious. Here we adopt the approximation
\be\label{eq_sigma_resolved}
	\sigma' = \sigma\,\sqrt{\frac{\alpha^2\cos i}{\gamma^2}},
\ee
where $\alpha$ is the angular half-mass diameter, measured along the major axis, of \ha~(for \ha~line) or \hm~(for CO lines), and $i$ is the galaxy inclination defined as the smaller angle between the line-of-sight and the galaxy's rotational axis. Our choice of using the the half-mass diameter rather than a larger diameter containing more of the gas mass relies on the assumption that the central concentration of gas can be exploited in clever algorithms for source extraction.

\subsection{Instantaneous bandwidth and spectral resolution}
ALMA's and SKA's limitations regarding the spectral resolution and the instantaneous bandwidth can be safely ignored within a study limited to the pure detection of extra-galactic emission lines in a narrow redshift range around $z=3$ (Sections \ref{section_detections_mw} and \ref{section_detections_general}). For typical observing times ($\Delta t\gg1/\Delta\nu$), the spectral resolution $\Delta\nu$ is mostly limited by the correlator performance. Both ALMA and SKA will be fitted with correlators allowing the selection of (frequency-dependent) spectral resolutions with an equivalent Doppler velocity far below $1\rm\,km\,s^{-1}$. Thus the chosen velocity channels of $75\rm\,km\,s^{-1}$ (see Section \ref{subsection_performace}) never conflict with the spectral resolution limit. As for the instantaneous spectral bandwidth (BW), ALMA ($\rm BW\leq\rm8\,GHz$) and SKA ($\rm BW\leq\min[0.25\,\nu_{\rm obs},\,\rm4\,GHz]$) are both able to cover a redshift range larger than $\Delta z=0.1$ at $z=3$, which is the range that will be considered in Section \ref{section_detections_general}. Even if the maximal instantaneous bandwidth were used, the implied highest spectral resolutions of ALMA (8192 channels) and SKA ($\sim\!10^4$ channels) still provide channels much smaller than $75\rm\,km\,s^{-1}$.

\subsection{Performance calculations for ALMA and SKA}\label{subsection_performace}

For all calculations hereafter frequency channels of an equivalent Doppler velocity of $75\rm\,km\,s^{-1}$ are assumed. Since ordinary star forming galaxies at $z\approx2$ show line widths of up to $\sim\!600\rm\,km\,s^{-1}$ (FWHM when seen edge-on, \citealp{Tacconi2010} and \citealp{Daddi2010}), channel widths larger than $75\rm\,km\,s^{-1}$ might be beneficial for the pure detection of cold gas at $z=3$. The predictions presented in this paper can be approximately rescaled to other channel widths $w$ by multiplying the signal-to-noise ratios $n$ (see definition in Section \ref{section_detections_mw}) by $\sqrt{w/(75\rm\,km\,s^{-1})}$. For example, a 3-$\sigma$ detection (i.~e.~$n=3$) with $75\rm\,km\,s^{-1}$ channels roughly corresponds to a 6-$\sigma$ detection (i.~e.~$n=6$) with $300\rm\,km\,s^{-1}$ channels. However, in practise the signal-to-noise ratio of large channels would be lowered by two mechanisms: the peaks in the line profile would get averaged-out, and a significant fraction of relatively face-on galaxies would have apparent line widths more narrow than the channel width.

Tab.~\ref{tab_sensitivities} displays the emission line specific performance of ALMA and SKA, as derived from the general telescope specifications in Tab.~\ref{tab_telescope_properties} and the equations introduced in the preceding part of this Section. The following list provides details and references for each column in Tab.~\ref{tab_sensitivities}:
\begin{enumerate}
  \item Emission line identifier.
  \item Telescope acronym. For SKA, the subscript specifies the construction phase, while LF (low frequency) and HF (high frequency) indicate the array type. For ALMA, the numbers indicate the frequency band.
  \item Rest-frame frequency of the line center.
  \item Observer-frame frequency of the line center, when observed at $z=3$.
  \item Channel width corresponding to an intrinsic velocity width (projected onto the line-of-sight) of $75\rm\,km\,s^{-1}$; calculated as $\Delta\nu=(75\rm\,km\,s^{-1}/c)\cdot\nu_{\rm obs}$.
  \item Receiver temperature. For ALMA these values are drawn from the current online specifications\footnote{http://www.eso.org/sci/facilities/alma/system/frontend/}; for SKA they are adopted from \cite{Dewdney2010}.\label{col_trec}
  \item Sky temperature.\label{col_tsky} For ALMA these values have been calculated as $T_{\rm sky}=T_{\rm b}\eta+T_{\rm atm}(1-\eta)$ [see appendix eq.~(1) in \citealp{Ishii2010}] with a source temperature $T_{\rm b}=T_{\rm CMB}=2.7\rm\,K$ (cosmic microwave background), an atmosphere temperature $T_{\rm atm}=260\rm\,K$, and atmospheric transparencies $\eta$ retrieved from ALMA's online$^5$ ``Atmospheric transmission calculator'' (for $\rm PVW=1.0$); for SKA-HF the sky temperature at 28.8\,GHz is considered equal to $T_{\rm CMB}$; for SKA-LF we follow \cite[][see Fig.~1 in \citealp{Dewdney2010}]{Bregman2000},
  \be\label{eq_tsky_HI}
  T_{\rm sky}=60{\rm\,K}\cdot(\lambda_{\rm obs}/{\rm m})^{2.55}.
  \ee
  \item System temperature $T_{\rm sys}=T_{\rm rec}+T_{\rm sky}$.
  \item ``Geometry factor'', representing the sampling efficiency of the wave front in the case of aperture arrays (see Section \ref{subsection_sensitivity}).
  \item ``Correlator quantization efficiency'', measuring the noise level added by the correlator (see Section \ref{subsection_sensitivity}).
  \item ``Array efficiency'', representing the sensitivity losses due to tapering (see Section \ref{subsection_sensitivity}).
  \item ``Antenna efficiency'' (also called ``aperture efficiency''), representing the energy fraction actually transferred from the collector to the receiver (see Section \ref{subsection_sensitivity}).
  \item Characteristic receiver noise $\sigma\,\sqrt{\Delta t}$ given by eq.~(\ref{eq_sigma}). Polarization information was ignored, i.~e.~$n_{\rm p}=2$, and the observations are assumed close enough to the zenith that $\cos(\theta)=1$ in eq.~(\ref{eq_aeff}).\label{col_point_source_sensitivity}
  \item Instantaneous \fov~calculated via eqs.~(\ref{eq_beamfov}, \ref{eq_instfov}). Note that the online ALMA specifications$^4$ use ${\rm FWHM}\approx\lambda_{\rm obs}/D$ instead of eq.~(\ref{eq_fwhm}), giving a \fov~30\% smaller than stated in Tab.~\ref{tab_sensitivities}.\label{col_fov}
  \item Spatial resolution calculated via eq.~(\ref{eq_gamma}).\label{col_spatial_resolution}
\end{enumerate}

\section{Detection of MW-type galaxies at $z=3$}\label{section_detections_mw}

Using the \sax~simulation (Section \ref{section_simulation}) and the telescope properties of ALMA and SKA (Section \ref{section_telescope_specifications}) we shall now investigate the ability of these telescopes to detect the redshifted CO and \ha~emission lines of galaxies at $z=3$. This section specifically addresses {\it MW-type} galaxies, while line detections in {\it arbitrary} galaxies will be considered in Section \ref{section_detections_general}.

From the wide range of possible observing goals at $z=3$ with ALMA and SKA we illustratively pick two questions: (i) What fraction of MW-type galaxies can be detected in each emission line at 3-$\sigma$ and 10-$\sigma$ significance in a 24\,h pointed observation? (ii) What observing time is required to detect a random single MW-type galaxy with 50\% chance? Here, ``observing time'' is defined as the integrated exposure time. It may, in practise, consist of multiple exposures spread over a period longer than the observing time itself.

We remind that (Section \ref{section_mw_evolution}) the 1731 simulated ``MW-type galaxies at $z=3$'' are defined as the most massive progenitors at $z=3$ (2.2\,Gyrs after the Big Bang) of all simulated galaxies that have MW-like properties (see Tab.~\ref{tab_mw_coldgas}) at $z=0$ (13.7\,Gyrs after the Big Bang). Therefore, the MW-type galaxies at $z=3$ represent a complete sample of MW progenitors within the semi-analytic model described in Section \ref{section_simulation}.

By definition, a galaxy will be called ``detected at $n$-$\sigma$'' in a particular emission line, if the peak flux density $s_{\rm p}$ of that line lies $n$-times above the RMS noise $\sigma$ of the observation, i.~e.~the ``signal-to-noise ratio'' is $n=s_{\rm p}/\sigma$. This is a conservative definition, since it makes no use of the spectral information. Combining multiple frequency channels (e.~g.~\citealp{Wang2006} for a related context) will undoubtedly increase the significance of a detection, although such sophisticated techniques still require some development.

The fraction of MW progenitors (all non-resolved, see Section \ref{subsection_spatial_resolution}) detected in \mbox{24\,h} at $n$-$\sigma$ can be obtained by integrating the normalized distribution of the peak flux densities $s_{\rm p}$ (lower panel of Fig.~\ref{fig_co_lum_distribution}) over $s_{\rm p}\geq n\cdot\sigma$, where $\sigma=(\sigma\sqrt{\Delta t})\,/\sqrt{(24\,{\rm h})}$ with $(\sigma\sqrt{\Delta t})$ drawn from Tab.~\ref{tab_sensitivities}, col.~(\ref{col_point_source_sensitivity}). On the other hand, the observing time $T$ required to detect 50\% of the MW progenitors at $n$-$\sigma$ is given by $T=(\sigma\,\sqrt{\Delta t})^2\,n^2\,\tilde{s}_p^2$, where $\tilde{s}_p$ is the median of the sample distribution of $s_{\rm p}$.

\begin{table*}[t]
\centering
\begin{tabular}{cccccccc}
\hline \\ [-2.0ex]
 ~~~~Emission~~~~ & ~~~~Telescope~~~~ & ~~~~$\log(\tilde{S})$~~~~ & ~~~~$\log(\tilde{s}_{\rm p})$~~~~ & \multicolumn{2}{c}{\mbox{~~~~Fraction of MW~~~~~~~~}} & \multicolumn{2}{c}{\mbox{~~~~~~~~Time to detect~~~~~~~~}} \\
 line & and band & $\rm[Wm^{-2}]$ & \multicolumn{1}{c}{$\rm[Jy]$} & \multicolumn{2}{c}{\mbox{~~~~detected in 24 h~~~~~~~~}} & \multicolumn{2}{c}{\mbox{50\% of MWs [h]}} \\
&&&& \multicolumn{1}{c}{3-$\sigma$} & \multicolumn{1}{c}{10-$\sigma$} & \multicolumn{1}{c}{3-$\sigma$} & \multicolumn{1}{c}{10-$\sigma$} \\
(1) & (2) & (16) & (17) & (18) & (19) & (20) & (21) \\ [0.5ex]
\hline \\ [-2.0ex]
\mbox{HI} & \mbox{SKA$_1$-LF} & \mbox{$-27.9\,/\,-27.2$} & \mbox{$-6.9\,/\,-6.2$} & \mbox{$0.0\%\,/\,0.0\%$} & \mbox{$0.0\%\,/\,0.0\%$} & \mbox{$2.8\!\cdot\! 10^{7}\,/\,1.1\!\cdot\! 10^{6}$} & \mbox{$3.1\!\cdot\! 10^{8}\,/\,1.2\!\cdot\! 10^{7}$} \\[0.5ex]
\mbox{HI} & \mbox{SKA$_2$-LF} & \mbox{$-27.9\,/\,-27.2$} & \mbox{$-6.9\,/\,-6.2$} & \mbox{$0.0\%\,/\,0.0\%$} & \mbox{$0.0\%\,/\,0.0\%$} & \mbox{$1.1\!\cdot\! 10^{6}\,/\,4.4\!\cdot\! 10^{4}$} & \mbox{$1.2\!\cdot\! 10^{7}\,/\,4.9\!\cdot\! 10^{5}$} \\[0.5ex]
\mbox{CO(1--0)} & \mbox{SKA$_1$-HF} & \mbox{$-24.0$} & \mbox{$-5.2$} & \mbox{$5.6\%$} & \mbox{$0.3\%$} & \mbox{$360$} & \mbox{$4.0\!\cdot\! 10^{3}$} \\[0.5ex]
\mbox{CO(1--0)} & \mbox{SKA$_2$-HF} & \mbox{$-24.0$} & \mbox{$-5.2$} & \mbox{$92.1\%$} & \mbox{$35.4\%$} & \mbox{$3.6$} & \mbox{$40$} \\[0.5ex]
\mbox{CO(3--2)} & \mbox{ALMA-3} & \mbox{$-22.6$} & \mbox{$-4.3$} & \mbox{$48.1\%$} & \mbox{$7.5\%$} & \mbox{$25$} & \mbox{$280$} \\[0.5ex]
\mbox{CO(4--3)} & \mbox{ALMA-3} & \mbox{$-22.4$} & \mbox{$-4.2$} & \mbox{$34.3\%$} & \mbox{$5.4\%$} & \mbox{$43$} & \mbox{$480$} \\[0.5ex]
\mbox{CO(5--4)} & \mbox{ALMA-4} & \mbox{$-22.4$} & \mbox{$-4.4$} & \mbox{$41.5\%$} & \mbox{$10.2\%$} & \mbox{$34$} & \mbox{$380$} \\[0.5ex]
\mbox{CO(6--5)} & \mbox{ALMA-5} & \mbox{$-22.8$} & \mbox{$-4.8$} & \mbox{$16.0\%$} & \mbox{$4.8\%$} & \mbox{$360$} & \mbox{$4.0\!\cdot\! 10^{3}$} \\[0.5ex]
\mbox{CO(7--6)} & \mbox{ALMA-5} & \mbox{$-23.4$} & \mbox{$-5.4$} & \mbox{$9.1\%$} & \mbox{$3.5\%$} & \mbox{$5.3\!\cdot\! 10^{3}$} & \mbox{$5.8\!\cdot\! 10^{4}$} \\[0.5ex]
\mbox{CO(8--7)} & \mbox{ALMA-6} & \mbox{$-24.1$} & \mbox{$-6.2$} & \mbox{$4.6\%$} & \mbox{$1.1\%$} & \mbox{$2.8\!\cdot\! 10^{5}$} & \mbox{$3.1\!\cdot\! 10^{6}$} \\[0.5ex]
\mbox{CO(9--8)} & \mbox{ALMA-6} & \mbox{$-25.0$} & \mbox{$-7.2$} & \mbox{$1.7\%$} & \mbox{$0.3\%$} & \mbox{$2.1\!\cdot\! 10^{7}$} & \mbox{$2.3\!\cdot\! 10^{8}$} \\[0.5ex]
\mbox{CO(10--9)} & \mbox{ALMA-7} & \mbox{$-26.1$} & \mbox{$-8.3$} & \mbox{$0.1\%$} & \mbox{$0.0\%$} & \mbox{$8.8\!\cdot\! 10^{9}$} & \mbox{$9.8\!\cdot\! 10^{10}$} \\[0.5ex]
\hline
\end{tabular}
   \caption{Simulated detectability of the redshifted emission lines emitted by MW progenitors at $z=3$. The analysis is based on a complete sample of 1731 simulated MW progenitors. $\tilde{S}$ and $\tilde{s}_{\rm p}$ denote sample medians of the integrated line flux and the peak flux density, respectively. Detailed descriptions of each column are provided at the beginning of Section \ref{section_detections_mw}. For \ha~detections two values are given, corresponding to the raw \sax~output ($\fudge=1$) and the corrected one ($\fudge=5$), as described in Section \ref{subsec_missing_ha}.}
   \label{tab_mw_detection}
\end{table*}

The results of these calculations are provided in Tab.~\ref{tab_mw_detection}. The columns have been numbered such that Tab.~\ref{tab_mw_detection} becomes an extension of Tab.~\ref{tab_sensitivities}, i.~e.~identical columns are given the same column number, while new columns are given a consecutive column number. These additional columns are:
\begin{enumerate}
\setcounter{enumi}{15}
  \item Median value of $\log(S/\rm{[W\,m^{-2}]})$ in the sample of 1731 simulated MW progenitors at $z=3$, where $S$ is the frequency-integrated line flux.\label{col_intflux}
  \item Median value of $\log(s_{\rm p}/\rm{Jy})$ in the sample of 1731 simulated MW progenitors at $z=3$, where~$s_{\rm p}$ is the peak flux density of the emission line.
  \item Fraction of simulated MW progenitors at $z=3$ detected at 3-$\sigma$ (or higher) in a 24\,h observation. Values below $0.1\%$ are not resolved here, since they correspond to less than one simulated galaxy.\label{col_fractionmw}
  \item Same as col.~(\ref{col_fractionmw}) but with 10-$\sigma$ detection limit.
  \item Exposure time required to detect a random simulated MW progenitor at $z=3$ at 3-$\sigma$ (or higher) with a chance of 50\%.\label{col_expmw}
  \item Same as col.~(\ref{col_expmw}) but with 10-$\sigma$ detection limit.
\end{enumerate}

The results in Tab.~\ref{tab_mw_detection} directly address the first ALMA science goal to detect spectral line emission from CO in a normal galaxy like the MW at a redshift of $z=3$, in less than 24\,h of observation. Col.~(\ref{col_fractionmw}) of Tab.~\ref{tab_mw_detection} reveals that only the lines between CO(3--2) and CO(6--5) inclusive can make a serious contribution to the number of detections in the sense that each of those lines will be detected at a 3-$\sigma$ level (or higher) in more than 10\% of all model scenarios for the MW at $z=3$. This fraction becomes maximal for the CO(3--2) line, contained in ALMA band 3. The same ALMA band also contains the redshifted CO(4--3) line, which has the second highest detection rate according to our predictions. {\it We therefore conclude that ALMA band 3 can best respond to the first ALMA science goal.} The odds for a MW detection with ALMA-3 can even be increased, assuming an ALMA correlator that allows the simultaneous observation of CO(3--2) and CO(4--3) at $z=3$. In this case, the signal-to-noise ratio $n$ of the co-added peak flux densities $s_{\rm p}$ (``single source stacking'') can be approximated as
\be
n=\frac{s_{\rm p,CO(3-2)}+s_{\rm p,CO(4-3)}}{\sqrt{\sigma^2_{\rm CO(3-2)}+\sigma^2_{\rm CO(4-3)}}}\,
\ee
We find that the odds of detecting a random MW progenitor at $z=3$ (3-$\sigma$ significance) are as high as 60\%, if both the CO(3--2) and CO(4--3) line are used simultaneously.

The higher order transitions including and above CO(7--6) will be virtually non-detectable. All those transitions require about or more than 1\,yr ($=8.76\cdot10^3\rm\,h$) of effective observing time to pick up a random MW progenitor with 50\% chance at 3-$\sigma$ significance. The few cases ($<10\%$) detected in 24\,h, correspond to the objects in the pronounced wings of the flux distributions shown in Fig.~\ref{fig_co_lum_distribution}. Those objects represent the simulation scenarios where the MW underwent a massive starburst or significant black hole activity exactly at a cosmological time corresponding to $z=3$ (see Section \ref{subsection_hacoz3}). Those cases are therefore outside the scope of the first ALMA goal to detect a ``normal'' galaxy like the MW.

CO(1--0) observations with the core of SKA-HF -- should it reach 28.8\,GHz -- will be very powerful. SKA$_2$-HF can detect more than 90\% of all model MWs at $z=3$ at 3-$\sigma$ significance (or more) in 24\,h. This high detection rate directly results from SKA's huge collecting area, which widely exceeds that of ALMA, even when accounting for the relatively small aperture efficiency of $\epsilon_{\rm a}=0.56$ at 28.8\,GHz.

On the other hand, SKA-LF will be virtually unable to detect \ha~emission from a MW progenitor at $z=3$, as can be seen from the very long observing times ($\gg1\rm\,yr$) required to pick up 50\% of all MW progenitors in the simulation. Quite surprisingly, the detection rate of general galaxies in a blind $1\rm\,yr$ \ha~survey with SKA$_2$-LF will nonetheless be higher than that of any conceivable CO line survey with ALMA and SKA-HF (see Section \ref{subsection_gen_1yr}).

\section{Detection of arbitrary galaxies at $z=3$}\label{section_detections_general}

This section expands the scope of Section \ref{section_detections_mw} towards line detections in {\it arbitrary} galaxies at $z=3$ using ALMA and SKA. The Sections \ref{subsection_gen_24hrs}--\ref{subsection_gen_stacking} successively address three selected questions: (i) How many galaxies per unit sky area and redshift will be detected at 3-$\sigma$ and 10-$\sigma$ in a 24\,h single pointing? (ii) How many galaxies in a redshift range $\Delta z=0.1$ around  $z=3$ will be detected during a 1\,yr survey? (iii) What is the significance of a stacked signal obtained in a 24\,h observation of a redshift range $\Delta z=0.1$ around $z=3$?

The answers to (ii) and (iii) depend on the number of galaxies inside the observed \fov~and thus require the apparent galaxy positions, unlike in Section \ref{section_detections_mw}, where only the apparent peak flux densities were needed. Therefore this section can be regarded as a prototypical application of the \sax-Sky simulation \citep{Obreschkow2009f}, where both radiative and geometric properties are exploited.

Throughout the whole section, a ``single pointing'' refers to an observation of a field fixed on the sky with a solid angle equal to the instantaneous \fov~of the respective observation. As in Section \ref{section_detections_mw}, such an observation will, in practise, consist of several exposures spread over a period longer than the total observing time of the pointing. Furthermore we maintain the definition that a simulated galaxy is ``detected'' in a particular emission line at $n$-$\sigma$, if the peak flux density $s_{\rm p}$ of this line lies $n$ times above the RMS Jy-noise $\sigma$ given by eq.~(\ref{eq_sigma}); respectively above $\sigma'$ given by eq.~(\ref{eq_sigma_resolved}) in the case of extended sources.

For clarity, the results for the questions (i)--(iii) have been collected in a single table (Tab.~\ref{tab_general_detection}), although those questions will be explained and discussed separately in Sections \ref{subsection_gen_24hrs}--\ref{subsection_gen_stacking}. The columns of Tab.~\ref{tab_general_detection} have again been numbered in such a way that this table becomes an extension of Tab.~\ref{tab_sensitivities} and Tab.~\ref{tab_mw_detection}, i.~e.~identical columns are given the same column number, while new columns are given a new column number. The new columns in Tab.~\ref{tab_general_detection} are specified as follows.
\begin{enumerate}
\setcounter{enumi}{21}
  \item Differential number of galaxies at $z=3$, detected per square degree and unit of
redshift at a 3-$\sigma$ level (or higher) in a 24\,h single pointing observation. Where no value is given (symbol `--'), no galaxy in the \sax-Sky simulation can be detected. \label{col_diffnbcount}
  \item Same as (\ref{col_diffnbcount}) but with 10-$\sigma$ detection limit.\label{col_diffnbcountb}
  \item Absolute number of galaxies in the range $z=2.95\!-\!3.05$ detected at a 3-$\sigma$ level (or higher) in a 1\,yr observation using a single pointing for aperture arrays (SKA-LF) and 365 distinct 24\,h pointings for dishes (ALMA, SKA-HF).\label{col_absnb}
  \item Same as (\ref{col_absnb}) but with 10-$\sigma$ detection limit.\label{col_absnbb}
  \item Number of galaxies with $M_{\rm B}<-22\rm\,mag$ and $z=2.95\!-\!3.05$ inside the instantaneous \fov. If this \fov~contains less than one object on average, the number is set to 1, since we always target known objects.
  \item Signal-to-noise $n$ of a 24\,h stacking experiment, where the emission lines of all galaxies with $M_{\rm B}<-22\rm\,mag$ and $z=2.95\!-\!3.05$ in the instantaneous \fov~are co-added.\label{col_stacking_significance}
\end{enumerate}

\begin{table*}
\centering
\begin{tabular}{cccccccc}
\hline \\ [-2.0ex]
 Emission & ~~~~~~~~Telescope~~~~~~~~ & \multicolumn{2}{c}{\mbox{~~~${\rm d}N/({\rm d}z\,{\rm d}A)$ in 24\,h~~~~~~~~}} & \multicolumn{2}{c}{\mbox{Nb.~of detections in 1\,yr~~~~}} & Nb.~of stacked & \mbox{Signal-to-noise $n$} \\
 line & \mbox{and band} & \multicolumn{1}{c}{3-$\sigma$} & \multicolumn{1}{c}{10-$\sigma$} & \multicolumn{1}{c}{3-$\sigma$} & \multicolumn{1}{c}{10-$\sigma$} & galaxies & \multicolumn{1}{c}{of a 24\,h stacking} \\
 (1) & (2) & (22) & (23) & (24) & (25) & (26) & (27) \\ [0.5ex]
\hline \\ [-2.0ex]
\mbox{HI} & \mbox{SKA$_1$-LF} & \mbox{--\,/\,--} & \mbox{--\,/\,--} & \mbox{$51$\,/\,$1.2\!\cdot\! 10^{4}$} & \mbox{--\,/\,340} & \mbox{$1.1\!\cdot\!10^4$} & \mbox{1\,/\,5} \\[0.5ex]
\mbox{HI} & \mbox{SKA$_2$-LF} & \mbox{--\,/\,51} & \mbox{--\,/\,--} & \mbox{$1.2\!\cdot\! 10^{5}$\,/\,$3.4\!\cdot\! 10^{6}$} & \mbox{$3.4\!\cdot\! 10^{3}$\,/\,$3.6\!\cdot\! 10^{5}$} & \mbox{$1.1\!\cdot\!10^5$} & \mbox{15\,/\,77} \\[0.5ex]
\mbox{CO(1--0)} & \mbox{SKA$_1$-HF} & \mbox{$1.1\!\cdot\! 10^{4}$} & \mbox{$1.5\!\cdot\! 10^{3}$} & \mbox{$740$} & \mbox{$100$} & \mbox{1} & \mbox{2} \\[0.5ex]
\mbox{CO(1--0)} & \mbox{SKA$_2$-HF} & \mbox{$1.4\!\cdot\! 10^{5}$} & \mbox{$4.1\!\cdot\! 10^{4}$} & \mbox{$9.1\!\cdot\! 10^{3}$} & \mbox{$2.8\!\cdot\! 10^{3}$} & \mbox{1} & \mbox{21} \\[0.5ex]
\mbox{CO(3--2)} & \mbox{ALMA-3} & \mbox{$5.2\!\cdot\! 10^{4}$} & \mbox{$1.4\!\cdot\! 10^{4}$} & \mbox{$610$} & \mbox{$160$} & \mbox{1} & \mbox{9} \\[0.5ex]
\mbox{CO(4--3)} & \mbox{ALMA-3} & \mbox{$3.8\!\cdot\! 10^{4}$} & \mbox{$1.0\!\cdot\! 10^{4}$} & \mbox{$250$} & \mbox{$69$} & \mbox{1} & \mbox{8} \\[0.5ex]
\mbox{CO(5--4)} & \mbox{ALMA-4} & \mbox{$4.2\!\cdot\! 10^{4}$} & \mbox{$1.4\!\cdot\! 10^{4}$} & \mbox{$180$} & \mbox{$58$} & \mbox{1} & \mbox{12} \\[0.5ex]
\mbox{CO(6--5)} & \mbox{ALMA-5} & \mbox{$1.7\!\cdot\! 10^{4}$} & \mbox{$7.6\!\cdot\! 10^{3}$} & \mbox{$50$} & \mbox{$22$} & \mbox{1} & \mbox{6} \\[0.5ex]
\mbox{CO(7--6)} & \mbox{ALMA-5} & \mbox{$1.0\!\cdot\! 10^{4}$} & \mbox{$5.7\!\cdot\! 10^{3}$} & \mbox{$22$} & \mbox{$12$} & \mbox{1} & \mbox{3} \\[0.5ex]
\mbox{CO(8--7)} & \mbox{ALMA-6} & \mbox{$6.5\!\cdot\! 10^{3}$} & \mbox{$3.3\!\cdot\! 10^{3}$} & \mbox{$11$} & \mbox{$5.4$} & \mbox{1} & \mbox{0.9} \\[0.5ex]
\mbox{CO(9--8)} & \mbox{ALMA-6} & \mbox{$4.0\!\cdot\! 10^{3}$} & \mbox{$1.7\!\cdot\! 10^{3}$} & \mbox{$5.2$} & \mbox{$2.3$} & \mbox{1} & \mbox{0.2} \\[0.5ex]
\mbox{CO(10--9)} & \mbox{ALMA-7} & \mbox{$1.5\!\cdot\! 10^{3}$} & \mbox{$500$} & \mbox{$1.6$} & \mbox{$5.3\!\cdot\! 10^{-1}$} & \mbox{1} & \mbox{0.03} \\[0.5ex]
\hline
\end{tabular}
   \caption{Simulated detectability of the redshifted emission lines of arbitrary galaxies at $z=3$. The 1\,yr number counts and the stacking analysis depend on the respective instantaneous \fov. Detailed descriptions of each column are provided at the beginning of Section \ref{section_detections_general}. For \ha~detections two values are given, corresponding to the raw \sax~output ($\fudge=1$) and the corrected one ($\fudge=5$), as described in Section \ref{subsec_missing_ha}.}
   \label{tab_general_detection}
\end{table*}

\subsection{Differential number of galaxy detections in 24\,h}\label{subsection_gen_24hrs}
How many galaxies per unit solid angle $A$ and redshift $z$ will ALMA and SKA detected in CO and \ha~emission at $z=3$ via a 24\,h-single pointing? Formally speaking, we are asking for the differential number count $\rho\equiv\rm{d}N/(\rm{d}z\,\rm{d}A)$ at $z=3$. Due to the discrete number of galaxies, $\rho$ is computed as $\Delta N/(\Delta z\,\Delta A)$, where $\Delta N$ is the integer number of galaxies detected within the redshift range $\Delta z$ (around $z=3$) and inside the solid angle $\Delta A$. $\rho$ is independent of $\Delta z$ and $\Delta A$ if three conditions are met: (1) $\Delta z$ is small enough that the differences in the luminosity distances and the effects of cosmic evolution can be neglected; (2) $\Delta A$ is large enough to suppress the effects of cosmic variance; (3) the volume spanned by $\Delta z$ and $\Delta A$ is large enough that the shot noise on $\Delta N$ can be neglected. We can approximately satisfy these criteria by considering the total volume of the \sax-Sky simulation contained within the narrow redshift range $z=2.95\!-\!3.05$. In this redshift range the luminosity distance varies by 4\% and the cosmic look-back time by 82\,Mpc within the cosmology of the simulation (see Section \ref{section_simulation}). The considered volume approximately contains $2.88\cdot10^6$ simulated galaxies and covers a solid angle of $\Delta A\approx37.2\,\rm deg^2$, which corresponds to the comoving surface area of $(500/h\rm\,Mpc)^2$ (box size of the Millennium simulation) that is large enough to suppress the effects of cosmic variance.

$\rho$ can now be computed for each emission line and telescope by counting the number of galaxies $\Delta N$ with peak flux densities $s_{\rm p}$ greater or equal to $n$-times the Jy-noise of a 24\,h-observation. For non-resolved sources, this noise $\sigma$ is calculated via eq.~(\ref{eq_sigma}) or, analogously, via $\sigma=(\sigma\sqrt{\Delta t})/\sqrt{24\rm\,h}$, where $\sigma\sqrt{\Delta t}$ is drawn from col.~(\ref{col_point_source_sensitivity}) in Tab.~\ref{tab_sensitivities}. For resolved sources, the galaxy-dependent noise $\sigma'$ of eq.~(\ref{eq_sigma_resolved}) needs to be adopted instead. The resulting values for $\rho=\Delta N/(\Delta z\,\Delta A)$ are given col.~(\ref{col_diffnbcount}) and (\ref{col_diffnbcountb}) of Tab.~\ref{tab_general_detection}. Note that these values can be computed directly via the SQL-interface of the \sax-Sky simulation\footnote{http://s-cubed.physics.ox.ac.uk/queries/new?sim=s3\_sax}. For example, to get the differential number $\rho$ of CO(3--2) detections using ALMA band 3 (value in Tab.~\ref{tab_general_detection}, col.~\ref{col_diffnbcount}, row 5) the following query can be executed.
\vspace{0.2cm}
\hrule
\vspace{0.2cm}
\noindent
\textsl{select count(*)/37.2/0.1}\\
\textsl{from galaxies\_line}\\
\textsl{where zapparent between 2.95 and 3.05}\\
\textsl{and cointflux\_3*columpeak$>$3*0.63e-3/sqrt(24*60)}
\vspace{0.2cm}
\hrule
\vspace{0.3cm}
\noindent Explanation: \textsl{37.2} is the \fov~$\Delta A$ in $\rm deg^2$ of the \sax-Sky simulation at $z=3$; \textsl{0.1} is the redshift interval $\Delta z$; \textsl{zapparent} is the apparent redshift of the galaxies including peculiar velocities; \textsl{2.95} and \textsl{3.05} are the minimal and maximal values of zapparent; \textsl{cointflux\_3*columpeak} is peak flux density of the CO(3--2) line in units of Jy; \textsl{3} is the significance level of the detection; \textsl{0.63e-3} is the value of $\sigma\sqrt{\Delta t}$ in units of $\rm Jy\sqrt{\rm min}$ (copied from Tab.~\ref{tab_sensitivities}, col.~\ref{col_point_source_sensitivity}, row 5); \textsl{24*60} is the number of minutes per day. Note that the values in Tab.~\ref{tab_general_detection} may differ by up to 30\% from those output by the above SQL query, since Tab.~\ref{tab_general_detection} also accounts for the signal-to-noise decrease in the case of extended galaxies (see eq.~\ref{eq_sigma_resolved}).

Given the differential number counts $\rho$, the absolute numbers of line detections in a 24\,h single pointing are obtained through multiplying the values of $\rho$ by the instantaneous \fov~and by the instantaneous redshift range, which is dictated by the instantaneous bandwidth. Hence cols.~(\ref{col_diffnbcount}) and (\ref{col_diffnbcountb}) in Tab.~\ref{tab_general_detection} cannot be used for a direct comparison of the detection rates of the different lines, unless the observed sky field and redshift range are smaller than (and hence not limited by) the instantaneous \fov~and redshift range. The latter case is met, for example, when observing a small galaxy group at $z=3$. In this case, the highest CO detection rates are achieved using CO(1--0) [SKA$_2$-HF], followed by CO(3--2) [ALMA-3], CO(5--4) [ALMA-4], and CO(4--3) [ALMA-3].

By comparison, \ha~detections within a similarly small sky field at $z=3$ using a 24\,h SKA observation seem virtually impossible. Not a single galaxy of the $2.88\cdot10^6$ objects in the \sax-Sky simulation at $z=2.95\!-\!3.05$ has a peak flux density above the 3-$\sigma$ detection limit of SKA$_2$-LF. In fact, the 3-$\sigma$ detection limit corresponds to an \ha~mass of about $3\cdot10^{11}\rm\,M_\odot$ (assuming an intrinsic line width of $300\rm\,km\,s^{-1}$), which is heavier than the largest \ha~mass ever observed in the local Universe (e.~g.~\ha~Parkes All-Sky Survey, \citealp{Meyer2004}). As we shall demonstrate in Section \ref{subsection_gen_1yr}, \ha~nonetheless wins over CO by an appreciable difference during long surveys (here 1\,yr), where the differences in the instantaneous \fov~become crucial. Furthermore, Section \ref{subsection_gen_stacking} reveals that even in 24\,h observations \ha~can still be detected at $z=3$ when using SKA in combination with parallel redshift surveys.

\subsection{Absolute number of galaxy detections in 1\,yr}\label{subsection_gen_1yr}

How many galaxies in a redshift range $\Delta z=0.1$ around  $z=3$ will be detected in CO and \ha~emission during a 1\,yr-survey? Such a survey can, for example, serve as a measurement of the angular power spectrum and of the comoving space densities $\Omegaha(z=3)$ and $\Omegahm(z=3)$.

For such measurements, the precise position of the redshift range $\Delta z=0.1$ around $z=3$ may have to be adjusted individually for each emission line to avoid the frequencies of radio-frequency interferences (RFIs) and atmospheric absorption lines. However, for this analysis we shall use the interval $z=2.95\!-\!3.05$ while neglecting RFIs and atmospheric absorptions lines. The narrow redshift range $\Delta z=0.1$ was chosen to isolate this number count analysis from the effects of distance variations and cosmic evolution within the survey volume. Note, however, that both ALMA and SKA are foreseen to yield instantaneous bandwidths corresponding to larger ranges in redshift, while easily maintaining our spectral resolution of $75\rm\,km\,s^{-1}$.

As revealed in Sections \ref{section_detections_mw} and \ref{subsection_gen_24hrs}, SKA-LF is predicted to detect virtually no \ha~at $z=3$ within 24\,h. On the other hand, both ALMA and SKA-HF will be able to detect a significant amount of CO. This apparent advantage of dish-based CO detections over aperture array-based \ha~detections nonetheless vanishes when the instantaneous \fov~becomes important. In fact, aperture array-based \ha~searches can perform very long exposures of the same, very large sky field, while dishes must take many shorter exposures to map a significant sky field in the same total observing time. As an example, we here assume that a 1\,yr galaxy survey at $z=3$ is performed using a 1\,yr single pointing for SKA-LF (\ha~line) and 365 single pointings of 24\,h each for ALMA and SKA-HF (CO lines). In the latter case, the number of detected galaxies is readily obtained by multiplying the differential number counts ${\rm d}N/({\rm d}z\,{\rm d}A)$ in Tab.~\ref{tab_general_detection} (cols.~\ref{col_diffnbcount}, \ref{col_diffnbcountb}) by 365, by $\Delta z=0.1$, and by the instantaneous \fov~listed in Tab.~\ref{tab_sensitivities} (col.~\ref{col_fov}). In the approximation of non-extended sources, the number of simulated \ha~detections in a 1\,yr single pointing with SKA-LF can be counted by executing a new SQL-query on the \sax-Sky database (cf.~Section \ref{subsection_gen_24hrs}). For example, for a blind \ha~search using SKA$_2$-LF (Tab.~\ref{tab_general_detection}, col.~24, row 2) the query for this approximation reads 
\vspace{0.2cm}
\hrule
\vspace{0.2cm}
\noindent
\textsl{select count(*)/37.2*410}\\
\textsl{from galaxies\_line}\\
\textsl{where zapparent between 2.95 and 3.05}\\
\textsl{and hiintflux*hilumpeak$>$3*0.31e-3/sqrt(24*60*365)}
\vspace{0.2cm}
\hrule
\vspace{0.3cm}
\noindent Explanation: \textsl{37.2} is the \fov~in $\rm deg^2$ of the \sax-Sky simulation at $z=3$; \textsl{410} is the \fov~in $\rm deg^2$ of SKA$_2$-LF; \textsl{zapparent} is the apparent redshift of the galaxies including peculiar velocities; \textsl{2.95} and \textsl{3.05} are the minimal and maximal values of zapparent; \textsl{hiintflux*hilumpeak} is peak flux density of the \ha~line in units of Jy; \textsl{3} is the significance level of the detection; \textsl{0.31e-3} is the value of $\sigma\sqrt{\Delta t}$ in units of $\rm Jy\sqrt{\rm min}$ (copied from Tab.~\ref{tab_sensitivities}, col.~\ref{col_point_source_sensitivity}, row 2); \textsl{24*60*365} is the number of minutes per year. Note, however, that some values in Tab.~\ref{tab_general_detection} differ significantly (factor $\sim5$) from those output by the above SQL query, since Tab.~\ref{tab_general_detection} also accounts for the signal-to-noise decrease in the case of extended galaxies (see eq.~\ref{eq_sigma_resolved}).

The results for the absolute number of line detections in the range $z=2.95\!-\!3.05$ are provided in Tab.~\ref{tab_general_detection} (cols.~\ref{col_absnb}, \ref{col_absnbb}). A comparison with the differential number counts (cols.~\ref{col_diffnbcount}, \ref{col_diffnbcountb}) highlights the tremendous advantage of aperture arrays. Their giant \fov~compared to dishes fully compensates the weakness of \ha~emission compared to CO emission (e.~g.~Tab.~\ref{tab_mw_detection}, col.~\ref{col_intflux}). We further emphasize that the instantaneous \fov~of SKA-LF is here limited by the computational power of the digital back-end. In principle, at least an instantaneous \fov~of $10^4\rm\,deg^2$ is conceivable (hemisphere above an elevation of $30\rm\,deg$), and it seems to be only a matter of time until the respective computational resources will become available. Therefore, \ha~surveys with SKA will ultimately be faster than any CO survey with ALMA and SKA.

\subsection{Line stacking at $z=3$ in 24 hours}\label{subsection_gen_stacking}
\begin{figure}
  \includegraphics[width=\columnwidth]{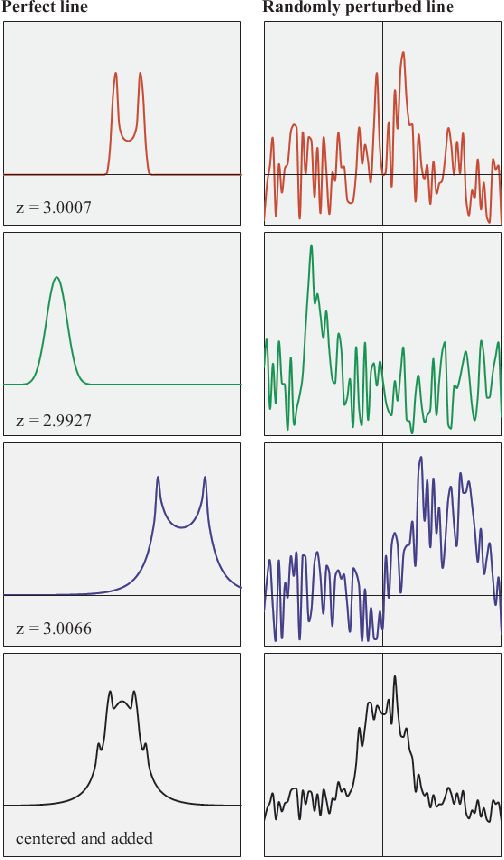}
  \caption{Idea of a line stacking experiment, illustrated with three simulated \ha~emission lines from different galaxies at $z\approx3$: a low-mass \ha-rich galaxy seen edge-on (red), a \ha-rich galaxy seen face-on (green), an intermediate mass galaxy seen edge-on (blue). Ideal simulated \ha~emission lines (left) are perturbed with Gaussian channel noise (right). Using the known redshift of each individual line (in reality drawn from spectroscopic optical/infrared redshift measurements), the lines are aligned and added to a single line (black). The stacked line yields an improved signal-to-noise, roughly a factor $\sqrt{3}$ above that of the individual lines.}
  \label{fig_stacking}
\end{figure}

What is the significance of a stacked signal obtained in a 24\,h survey of a redshift range $\Delta z=0.1$ around  $z=3$? Here ``stacking'' refers to the addition of possibly non-detected emission lines by using the positions and redshifts of their sources drawn from parallel surveys (e.~g.~optical/infrared). Stacking $N$ lines of comparable signal and noise ideally increases the signal-to-noise ratio by $\sqrt{N}$. It may hence become possible to measure the summed flux of otherwise non-detected emission lines. This technique is illustrated in Fig.~\ref{fig_stacking} for three random \ha~emission lines drawn from the \sax-Sky simulation. Stacking typically becomes useful for large samples of non-detected lines with redshift uncertainties $\delta z$ much smaller than the redshift interval spanned by an individual line, i.~e.~$\delta z<10^{-3}$ at $z=3$ (Fig.~\ref{fig_stacking}). For simplicity, we here neglected redshift and position errors, although they may be a major problem in real stacking experiments.

As shown in Fig.~\ref{fig_stacking}, a stacked line profile differs from those of single emission lines. The maximum flux density of a stacked line approximately corresponds to its central flux density, since all individual emission lines contribute to its central flux, but only the broader lines contribute to its tails. We therefore call a stacked line ``detected'' if its central flux density is detected. Formally, a stacked line composed of $N$ individual lines is detected at $n$-$\sigma$ if
\be\label{eq_stacking}
n=\frac{\sum_{i=1}^N{s_0^{\rm i}}}{\sqrt{N}\sigma},
\ee
where $s_0^{\rm i}$ is the central flux density of the $i$-th line and $\sigma$ is the channel noise (here assumed constant) given by eq.~(\ref{eq_sigma}) and Tab.~\ref{tab_sensitivities} (col.~\ref{col_point_source_sensitivity}).

Since stacking experiments require spectroscopic redshift measurements from parallel surveys, the latter impose sample selection criteria. We here consider a hypothetical stacking experiment using spectroscopic data from a galaxy survey limited to extinction-corrected absolute $B$-band magnitudes $M_{\rm B}<-22$. This sample definition approximately matches that of the Lyman-break galaxies (LBGs) at $z\approx3$, for which spectroscopic redshifts were obtained via optical follow-up observations at Keck I and II \citep{Steidel2003}. We stack the 24\,h emission line signals of all simulated galaxies with $M_{\rm B}<-22$ and $z=2.95\!-\!3.05$, contained in the instantaneous \fov~of the respective line (see Tab.~\ref{tab_sensitivities}, col.~\ref{col_point_source_sensitivity}). 

For the CO line observations (ALMA, SKA-HF), the instantaneous \fov~contains, on average, less than one such galaxy. In other words, stacking will not allow us to increase the signal-to-noise ratio. Only for comparison to \ha, we therefore say that the number of stacked galaxies is $N=1$ and we assume that this galaxy has CO fluxes corresponding to the geometric average of all simulated galaxies with $M_{\rm B}<-22$ and $z=2.95\!-\!3.05$. For \ha~line observations (SKA-LF), the instantaneous \fov~is very large; i.~e.~$41\rm\,deg^2$ (SKA$_1$-LF) and $410\rm\,deg^2$ (SKA$_2$-LF). The requirement of a spectroscopic redshift survey at $z=3$ covering such large \fov s, lies far beyond current possibilities (e.~g.~\citealp{Steidel2003} considered at total sky area of $0.38\rm\,deg^2$). Powerful wide-field spectrographs are needed to retrieve spectroscopic data on sky areas comparable to the instantaneous \fov~of the SKA-LF. Even the \sax-Sky simulation `only' covers a \fov~of $37.2\rm\,deg^2$ at $z=3$. In this stacking analysis we therefore linearly extrapolate the number of galaxies and their summed \ha~fluxes to the SKA \fov. 

The number of stacked galaxies $N$ and the signal-to-noise ratio $n$ of the stacked emission line can be computed using the online SQL-interface of the \sax-Sky database (cf.~Section \ref{subsection_gen_24hrs}). For example, for the \ha~line observed with SKA$_2$-LF (Tab.~\ref{tab_general_detection}, cols.~26, 27, row 2), the respective query reads
\vspace{0.2cm}
\hrule
\vspace{0.2cm}
\noindent
\textsl{select count(*)/37.2*410 as N,}\\
\textsl{sum(tab1.hiintflux*tab1.hilumcenter)/(0.31e-3/sqrt(24*60))/sqrt(count(*))*sqrt(410/37.2) as n}\\
\textsl{from galaxies\_line tab1, galaxies\_delu tab2}\\
\textsl{where tab1.id=tab2.id}\\
\textsl{and tab1.zapparent between 2.95 and 3.05}\\
\textsl{and tab2.mag\_bdust$<$-22}
\vspace{0.2cm}
\hrule
\vspace{0.3cm}
\noindent Explanation: \textsl{37.2} and \textsl{410} respectively denote the \fov s of the simulation and the survey; \textsl{tab1.hiintflux*tab1.hilumpeak} is peak flux density of the \ha~line in units of Jy; \textsl{0.31e-3} is the value of $\sigma\sqrt{\Delta t}$ in units of $\rm Jy\sqrt{\rm min}$ (copied from Tab.~\ref{tab_sensitivities}, col.~\ref{col_point_source_sensitivity}, row 2); \textsl{24*60} is the number of minutes per day; \textsl{2.95} and \textsl{3.05} are the minimal and maximal values of the redshift \textsl{tab1.zapparent}. This query also calls the table \textsl{galaxies\_delu} (\textsl{tab2}), which contains the extinction-corrected absolute $B$-band magnitudes \textsl{mag\_bdust} calculated in the semi-analytic galaxy model \citep{Croton2006,DeLucia2007}.

The significance levels $n$ of the stacked line detections, calculated via eq.~(\ref{eq_stacking}), are listed in Tab.~\ref{tab_general_detection} (col.~\ref{col_stacking_significance}). This analysis demonstrates that SKA$_2$-LF can reliably detect \ha~at $z=3$ in a 24\,h pointing using stacking techniques, given large and deep spectroscopic redshift surveys. This reveals again the strength of the large \fov~offered by aperture arrays. It should be emphasized that our $B$-band selection criterion ($M_{\rm B}<-22$) excludes most of the \ha~and CO at $z=3$. In fact, the stacked \ha~line only traces $2\%$ of the total \ha~mass in all simulated galaxies in the \fov. Larger fractions of the \ha~mass may be studiable by stacking on objects selected by star-formation rate indicators, such as used in the wide-area emission-line surveys planned with HETDEX \citep{Hill2004}.

\section{Discussion and conclusion}\label{section_conclusion}

\subsection{CO detections at $z=3$ with ALMA}
Will ALMA meet its primary science goal to detect MW-type galaxies at $z=3$? Yes it will. Just about. Beginning with a semi-analytic model, we selected 1928 simulated galaxies that resemble the MW at $z=0$ and backtracked their cosmic history to a time corresponding to $z=3$. In the resulting sample of ``MW progenitors'' or ``MW-type galaxies'' at $z=3$, ALMA has roughly a $50\%$ chance of detecting a random object in CO(3--2) emission at a 3-$\sigma$ level in a 24\,h pointing. ALMA band 3 is the best choice to achieve this goal, since it contains the CO(3--2) line and since it is the only band containing another low order CO transition at $z=3$, i.~e.~CO(4--3). If the instantaneous bandwidth is split into two windows, covering CO(3--2) and CO(4--3), and if these two lines are co-added (``single source stacking'', see Section \ref{section_detections_mw}), the odds of detecting a MW in 24\,h can be increased to 60\%. These predictions remain similar if instead of using a model for MW-progenitors at $z=3$, we simply imagined the actual MW at a cosmological distance equivalent to $z=3$, because the total \hm~mass only changes by about 20\% between the two cases (see Tab.~\ref{tab_mw_coldgas}). Whether those predictions will be met significantly depends on the final, currently uncertain sensitivity of the ALMA receivers and on the actual transparency of the atmosphere at the different bands.

As expected from its small instantaneous \fov, ALMA is a relatively slow survey instrument. On average, only about 1-2 general galaxies (not just MWs) per day will be detected in a blind CO survey between $z=2.95$ and $z=3.05$ (from col.~24 in Tab.~\ref{tab_general_detection}). To make effective use of ALMA as a CO survey instrument, it is hence crucial to preselect a sample using CO indicators, such as tracers of star formation.

\subsection{CO(1--0) and \ha~detections at $z=3$ with SKA}\label{subsec_discussion_ska}
SKA-HF -- should its frequency domain be extended up to 28.8\,GHz -- will provide a unique way of detecting CO at $z=3$. In fact SKA-HF searches for CO(1--0) promise to become much more effective in terms of number of detected objects than ALMA searches for CO(3--2). This result is based on the assumption that only the core of SKA-HF is used to keep most CO(1--0) sources non-resolved, and it already accounts for the fact that SKA-HF yields a low antenna efficiency $\epsilon_{\rm a}$ (see col.~12, Tab.~\ref{tab_sensitivities}) at 28.8\,GHz due to its limited dish surface accuracy. The power of SKA-HF compared to ALMA relies in two main reasons: firstly, SKA-HF will ultimately reach a total collecting area three orders of magnitude larger than that of ALMA, and secondly the instantaneous \fov~(i.e.~the beam size) is an order of magnitude larger for CO(1--0) than for CO(3--2).

On the other hand, \ha~detections at $z=3$ using SKA-LF are a mixed blessing. On the downside, Section~\ref{section_detections_mw} revealed that SKA-LF will be virtually unable to pick up a MW progenitor at $z=3$. Even when using the full SKA$_2$-LF array and when assuming that the typical MW progenitor yields 5-times more \ha~than predicted by the \sax~model, it would still take 5 years ($4.4\cdot10^4\rm\,h$, see col.~20, Tab.~\ref{tab_mw_detection}) of effective observing time to pick up such a MW progenitor at only $3$-sigma significance. We can conclude that MW progenitor studies at $z=3$ in \ha~will remain virtually impossible using the SKA, while CO detections with ALMA seem possible within 24\,h. On the other hand, Section~\ref{section_detections_general} suggests that blind searches for \ha~in \textit{general} galaxies (not just MW progenitors) at $z=3$ will be more effective than any conceivable CO survey with ALMA and SKA-HF. In fact, the huge instantaneous \fov~of SKA-LF, which is roughly 6 orders of magnitude larger than that of ALMA, compensates for the low fluxes of \ha~lines compared to those of CO lines. As a result SKA$_2$-LF promises to detect above $10^5$, perhaps even above $10^6$, galaxies at $z=2.95-3.05$ in a 1 year blind search for \ha, while ALMA will find less than $10^3$ CO emitters in the same time. Furthermore, if the positions and redshifts of the most \ha-rich galaxies in the SKA-LF \fov~are already known from a parallel spectroscopic survey, then stacking techniques can be applied to detect \ha~in less than 24\,h (col.~27, Tab.~\ref{tab_general_detection}).

Finally, it is worth considering the implications of extending the mid-frequency array SKA-MF down to 355\,MHz to observe the \ha~line at $z=3$. In this case, the wave front of the \ha~line can be sampled completely, implying a geometry factor close to $\epsilon_g=1$ compared to $\epsilon_g=0.1$ of SKA-LF (see Tab.~\ref{tab_sensitivities}). The implied gain in sensitivity is roughly compensated by the 10-times smaller collecting area of SKA-MF compared to SKA$_2$-LF. However, SKA-MF would still have better point source sensitivity because of its lower receiver temperatures ($T_{\rm rec}\approx50\rm\,K$, see Tab.~\ref{tab_telescope_properties}) and because it will provide a better image quality due to reduced side-lobes above the horizon. The instantaneous \fov~is also likely to be significantly larger due to the smaller station size. In conclusion, SKA-MF promises interesting features for observing \ha~at $z\approx3$.

\subsection{Closing words}
This joint study of ALMA and SKA has highlighted remarkable differences in their performances and applicabilities. In general, ALMA is most powerful at targeted observations, while SKA will be very suitable for long blind searches, thus placing ALMA preferentially in the surroundings of galaxy evolution, while SKA will reach deep into large scale cosmology. Thus ALMA and SKA will be heavily synergetic. Together those instruments will set a supreme long-term standard in speed and sensitivity over the whole atmospherically transparent frequency domain between 70\,MHz and 920\,GHz; and together they will resolve many outstanding questions across all redshifts of the star-forming universe.

With minor exceptions, this paper has been restricted to $z=3$ and to the current benchmarks for ALMA and SKA. However, the methods and on-line simulation tools presented here can be applied to any other redshift and any other telescope. This paper therefore sets the stage for the statistical comparison of future cold gas surveys with simulations. Such a comparison is a crucial, if not the only, way to verify physical theories of galaxy formation against the empirical reality.

\section{Acknowledgements}
We thank Andrew Baker for vivid discussions and the anonymous referee for inspiring inputs. The Millennium Simulation databases used in this paper and the web application providing online access to them were constructed as part of the activities of the German Astrophysical Virtual Observatory. DO thanks Dr.~Aris Karastergiou for representing Eastside.


\end{document}